\documentclass[twocolumn,english]{revtex4-1}
\usepackage[LGR,T2A,T1]{fontenc}
\usepackage[latin9]{inputenc}
\usepackage[a4paper]{geometry}
\usepackage{color}
\usepackage{array}
\usepackage{multirow}
\usepackage{amsmath}
\usepackage{stmaryrd}
\usepackage{graphicx}

\makeatletter

\DeclareRobustCommand{\greektext}{%
  \fontencoding{LGR}\selectfont\def\encodingdefault{LGR}}
\DeclareRobustCommand{\textgreek}[1]{\leavevmode{\greektext #1}}
\DeclareFontEncoding{LGR}{}{}
\DeclareTextSymbol{\~}{LGR}{126}
\DeclareRobustCommand{\cyrtext}{%
  \fontencoding{T2A}\selectfont\def\encodingdefault{T2A}}
\DeclareRobustCommand{\textcyr}[1]{\leavevmode{\cyrtext #1}}
\AtBeginDocument{\DeclareFontEncoding{T2A}{}{}}

\providecommand{\tabularnewline}{\\}
\newcommand{\lyxdot}{.}

\makeatother

\usepackage{babel}
\begin{document}

\title{Lattice distortions and/or intercalation as ways to induce magnetism
in \textgreek{a}-FeSi$_2$: a theoretical study. }

\author{V.Zhandun$^{1}$},
\email{jvc@iph.krasn.ru}
\author{N. Zamkova$^{1}$, P. Korzhavyi$^{2}$, I. Sandalov$^{1}$}

\affiliation{$^{1}$Kirensky Institute of Physics, Federal Research Center \textquotedbl{}Krasnoyarsk
Science Centre, Siberian Branch of the Russian Academy of Sciences'',
660036 Krasnoyarsk, Russia; $^{2}$Kungliga Tekniska H$\ddot{o}$gskolan, SE-100
44 Stockholm, Sweden}
\begin{abstract}
The possibilities to induce magnetism in the non-magnetic bulk $\alpha-\mathrm{FeSi}{}_{2}$
by means of lattice distortions or intercalation with metal or non-metal
ions of light elements is investigated theoretically by combined \emph{ab
initio} and model methods. We find that the distortions indeed can
induce the formation of magnetic moment on iron atoms in certain local
environments; however, the required strength of the distortions often
is too large to be achieved in experiments. For this reason we suggest
using ``chemical pressure'' that is, intercalating the $\alpha-\mathrm{FeSi}{}_{2}$
films by light elements. We find that some of such variants have promising
characteristic.
\end{abstract}
\maketitle

\section{Introduction. }

The modern semiconductor industry is mainly based on the silicon \cite{key-1}.
The spintronics development demands for new magnetic materials compatible
with silicon. These facts motivate for a search of the transition
metal silicides which are either magnetic or close to the magnetic
instability. The ability of iron to form a vast variety of magnetic
compounds with silicon both in the bulk and in the epitaxially stabilized
forms makes them especially attractive. These compounds are already
used in micro- and optoelectronics, an also in photovoltaics \cite{key-2}-\cite{key-6}.\textbf{
}Iron disilicide\textbf{ $\alpha-\mathrm{FeSi}{}_{2}$} is unstable
and non-magnetic in the bulk form. For these reasons it was not in
the first lines on the list of candidates for applications.
The situation has changed after publications
\cite{key-7,key-8,key-9} where it was shown that the film and nanoparticles
of $\alpha-\mathrm{FeSi}{}_{2}$ can be epitaxially stabilized. Moreover,
it becomes magnetic. These experimental achievements have good perspective
for the integration of the FeSi-based magnetic devices into silicon
technology. However, a sensible choice of the optimal technology has
to be based on a detailed understanding of the physics of the magnetic
moment formation in these compounds, which
is not achieved yet. One of the factors leading to the appearance
of magnetism in these compounds can be lattice distortions. Particularly,
for $\alpha-\mathrm{FeSi}{}_{2}$  the experimental data \cite{key-7,key-8,key-9}
and theoretical analysis \cite{key-10} show that they are essential
ingredients for the formation of magnetism both in films \cite{key-7}
and nano-particles \cite{key-8,key-9}.

The reason why certain lattice distortions can favor the magnetism formation
is seen from the second order of perturbation theory. Itinerant magnetism arises due
to peaks in the density of electron states. These peaks originate
either from the presence of narrow bands, or from flat areas on the
Fermi surface. If a band has a large bandwidth and does not contain
narrow enough peaks in DOS in the vicinity of Fermi energy, a Stoner-like
criterium for magnetism is not fulfilled and magnetic moment (MM)
is not formed. Therefore, any mechanism which favors a decrease of
width of an effective d-band, will favor also magnetism formation.
If a d-electron may hop to a neighboring atom (\emph{nAt}), and the
atomic levels $\varepsilon_{Fe-Fe}^{0}$ and $\varepsilon_{nAt-nAt}^{0}$
are separated, then its band is renormalized roughly as follows:

\[
\varepsilon_{Fe-Fe}^{*}(k)=\varepsilon_{Fe-Fe}(k)+\frac{\left|t_{Fe-nAt}(k)\right|^{2}}{\varepsilon_{Fe-Fe}(k)-\varepsilon_{nAt-nAt}(k)}.
\]
Then an increase\textbf{ }of the distance between \emph{Fe} atom and
\emph{nAt} decreases the hopping matrix element $t_{Fe-nAt}(k)$ and,
therefore, the effective width of the \emph{d}-band decreases too
and makes the fulfillment of the Stoner's criterium easier.

Earlier discussions were focussed on the effect of in-plane distortions
caused by the misfit strains \cite{key-10}. However, according to
recent experimental data \cite{key-11} besides the in-plane distortions,
out-of-plane distortions also can arise in $\alpha-\mathrm{FeSi}{}_{2}$
nanoparticles.\textbf{ }The observed magnetic moments are quite small
\cite{key-11,key-7} $\sim0.2\mu_{B}$. Theoretical analysis of the
magnetism formation in iron silicides \cite{key-10} shows that the
small lattice distortions which arise during the film fabrication
may cause small moments. One can expect that the increase of these
distortions may lead to an increase of the magnetic moments. Since
the crystal structure of iron disilicide $\alpha-\mathrm{FeSi}{}_{2}$ has
a cavity formed by Si planes, one way to increase these distortions
is intercalation of $\alpha-\mathrm{FeSi}{}_{2}$ with light atoms.

The present work is devoted to further theoretical analysis of the
mechanisms of magnetism formation in the disilicide of iron $\alpha-\mathrm{FeSi}{}_{2}$.
We inspect the possibilities to induce magnetism by means of ``chemical
pressure'' via intercalation the $\alpha-\mathrm{FeSi}{}_{2}$ by
different light elements. We use \emph{ab initio} (VASP, DFT-GGA,
see Sec. \ref{sec:Calculation-details.}) approach along with a hybrid
approach\emph{ }that combines \emph{ab initio} and model calculations,
developed in Refs. \cite{key-10,key-12}. Then we map the DFT-GGA
results onto the multiorbital model, suggested in Ref. \cite{key-12}.
The mapping is based on the idea to exploit the Hohenberg-Kohn theorem,
equalizing the charge densities, one generated by the Kohn-Sham equations
and obtained from the Hartree-Fock equations for a model Hamiltonian.
Due to success of the Kohn-Sham approach in description of real materials
we treat the corresponding charge density as a ``genuine'' one and
find the parameters of the model Hamiltonian from minimization of
difference beween the Kohn-Sham and the model Hartree-Fock charge
densities.

The analysis of the model allows for detailed understanding
of the role played by different parameters of the model in the physics
of magnetism formation.

The paper organised as follows. In Sec. II, we provide
the details of the {\emph{ab initio}}
and the model calculations. The effect of the lattice distortions
in $\alpha-\mathrm{FeSi}_{2}$ on magnetic moment
formation in both approaches is described in Sec. III A. The Sec.
III B presents the results of the calculations for $\alpha-\mathrm{FeSi}{}_{2}$
with intercalated atoms. Section IV contains the conclusions.

\section{Calculation details.\label{sec:Calculation-details.}}

\subsection{\emph{Ab initio} part}

All presented here \emph{ab initio} calculations have been performed
using the Vienna \emph{ab initio} simulation package (VASP) \cite{key-13}
with projector augmented wave (PAW) pseudopotentials \cite{key-14}.
The valence electron configuration $3d^{6}4s^{2}$ is taken for the
Fe atoms and the $3s^{2}3p^{2}$ one for the Si atoms. The calculations
are based on density functional theory (DFT) in the generalized gradient
approximation (GGA), where the exchange-correlation functional is
chosen within the Perdew-Burke-Ernzerhoff (PBE) parametrization \cite{key-15}.
Throughout all calculations, the plane-wave cutoff energy was 500
eV, and the Gauss broadening with a smearing of 0.05 eV was used.
The Brillouin-zone integration was performed on a $15\times15\times8$
Monkhorst-Pack grid \cite{key-16} of special points. The optimized
lattice parameters and atomic coordinates were obtained by minimizing
the total energy.

\subsection{Model part}

In \cite{key-12} we suggested to combine the \emph{ab initio} and
model calculations by means of the following scheme. First, we perform
\emph{ab initio} calculations of electronic and magnetic properties
within the framework of DFT-GGA. Then we map the DFT-GGA results onto
the multiorbital model suggested in Ref. \cite{key-12}. The details
of model calculations are described in Ref. \cite{key-12}. Here we
give only the Hamiltonian and the general parameters of the model.
We use the set of the Kanamori interactions \cite{key-17} between
the $d$-electrons of Fe (five $d$-orbitals per spin). The crystal
structure contains neighboring Fe ions, for this reason the direct
interatomic $d-d$-exchange and $d-d$-hopping have to be included.
The Si $p$-electrons (three $p$-orbitals per spin) are modeled by
atomic levels and interatomic hoppings. Both subsystems are connected
via $d-p$-hoppings. Thus, the Hamiltonian of the model is:

\begin{widetext}
\begin{equation}
H=H^{Fe}+H_{J'}^{Fe-Fe}+H_{0}^{Si}+H_{hop},
\end{equation}
where

\[
H^{Fe}=H_{0}^{Fe}+H_{K}^{Fe}
\]

\begin{center}
$H_{0}^{Fe}=\sum\varepsilon_{0}^{Fe}\hat{n}_{nm\sigma}^{d}.$
\par\end{center}

The Kanamori\textquoteright s part of the Hamiltonian is

\begin{multline}
H_{K}^{Fe}=\frac{U}{2}\sum\hat{n}_{nm\sigma}^{d}\hat{n}_{nm\bar{\sigma}}^{d}+\left(U'-\frac{1}{2}J\right)\sum\hat{n}_{nm}^{d}\hat{n}_{nm'}^{d}\left(1-\delta_{mm'}\right)-\frac{1}{2}J\sum\hat{\boldsymbol{s}}_{nm}^{d}\hat{\boldsymbol{s}}_{nm'}^{d}.
\end{multline}

The Hamiltonian of the interatomic exchange and hopping parts is

\begin{multline}
H_{J'}^{Fe-Fe}=-\frac{1}{2}J'\sum\hat{\boldsymbol{s}}_{nm}^{d}\hat{\boldsymbol{s}}_{n'm'}^{d};\\
H_{hop}=\sum T_{n,n'}^{mm'}p_{nm\sigma}^{\dagger}p_{n'm'\sigma}+\sum t_{n,n'}^{mm'}d_{nm\sigma}^{\dagger}d_{n'm'\sigma}+\sum\left[\left(t'\right)_{n,n'}^{mm'}d_{nm\sigma}^{\dagger}p_{n'm'\sigma}+H.c.\right];
\end{multline}

where

\begin{multline}
\hat{n}_{nm\sigma}^{d}\equiv d_{nm\sigma}^{\dagger}d_{nm\sigma};\;\hat{n}_{nm}^{d}=\hat{n}_{nm\uparrow}^{d}+\hat{n}_{nm\downarrow}^{d};\;\hat{\boldsymbol{s}}_{nm}^{d}\equiv\boldsymbol{s}_{\alpha\gamma}d_{nm\alpha}^{\dagger}d_{nm\gamma};\;\hat{n}_{nm\sigma}^{p}\equiv p_{nm\sigma}^{\dagger}p_{nm\sigma}.
\end{multline}

\end{widetext}

Here $p^{\dagger}$ ($p$) are the creation (annihilation) operators
of $p$-electrons of Si and $d^{\dagger}$ and $d$ stand for $d$-electrons
of Fe ions; $n$ is the complex lattice index (site, basis); $m$
labels the orbitals; the indices $\sigma,\alpha,\gamma$ are spin
projections; $\boldsymbol{s}$ are Pauli matrices; $U,\:U'$ and $J$
are the intra-atomic Kanamori parameters; $J'$ is the parameter of
the intersite exchange between the nearest Fe atoms. At last, $T_{n,n'}^{mm'},\;t_{n,n'}^{mm'}\;\left(t'\right)_{n,n'}^{mm'}$
are hopping integrals between Si - Si, Fe -Fe and Fe - Si atomic pairs,
correspondingly.

The dependencies of hopping integrals on $k$ were obtained from the
Slater and Koster atomic orbital scheme \cite{key-18} in the two-center
approximation using a basis set consisting of five $3d$ orbitals
for each spin on each Fe atom and three $3p$ orbital for each spin
on each Si atom. Then, within the two-center approximation, the hopping
integrals are expressed in terms of the Slater \textendash{} Koster
parameters $t_{\sigma}\equiv(dd\sigma)$, $t_{\pi}\equiv(dd\pi)$
and $t_{\delta}\equiv(dd\delta)$ for Fe - Fe hopping and $t_{\sigma}\equiv(pd\sigma)$,
$t_{\pi}\equiv(pd\pi)$ for Fe - Si  and Si - Si hoppings. In calculations
of the model phase diagrams (maps) for magnetic moments we neglected
the weak $\delta-$ bonds $(t_{\delta}=0)$ for Fe
- Fe hopping and kept fixed the relations $t_{\pi}=t_{\sigma}/3$
for the nearest neighbors (NN) Fe -Si ($t_{\sigma}\equiv t_{Fe-Si}$)
and $t_{\pi}=t_{\sigma}/2$ for the next nearest neighbors (NNN) Fe
- Fe ($t_{\sigma}\equiv t_{Fe-Fe}$) and Si - Si ($t_{\sigma}\equiv t_{Si-Si}$)
and $t_{\pi}=t_{\sigma}/2$. \textcolor{black}{We assume that hopping
integrals depend on the distance }\textcolor{black}{\emph{$R$}}\textcolor{black}{{}
between the ions exponentially, }

\begin{equation}
t(R)=t^{max}exp(-\gamma\Delta R)
\end{equation}

\textcolor{black}{where }$t^{max}=t(R_{min})$\textcolor{black}{{} and
}$\Delta R=R-R_{min}\mathrm{(\mathring{A}})$\textcolor{black}{. We
have found the parameters}\textbf{\textcolor{black}{{} }}$\gamma_{1}=0.89\mathrm{\mathring{A}}^{-1}$\textbf{\textcolor{black}{{}
}}\textcolor{black}{for $t_{Fe-Fe}$}\textbf{\textcolor{black}{, }}$\gamma_{2}=0.93\mathrm{\mathring{A}}^{-1}$\textbf{\textcolor{black}{)
}}\textcolor{black}{for $t_{Fe-Si}$ \cite{key-10} and}\textbf{\textcolor{black}{{}
}}$\gamma_{3}=0.94\mathrm{\mathring{A}}^{-1}$\textbf{\textcolor{black}{{}
}}\textcolor{black}{for}\textbf{\textcolor{black}{{} }}\textcolor{black}{$t_{Si-Si}$.
}\textit{\textcolor{black}{\emph{The on-site parameters}}} during
all model calculations were the following: $U=1\:\mathrm{eV}$\emph{,}
$J=0.4\:\mathrm{eV},\:\varepsilon_{Si}=6\:\mathrm{eV}\:\varepsilon_{Fe}=0$
. In the rest of the paper all hopping parameters are given in eV.

\section{Results and discussion }

\subsection{The effect of lattice distortions on magnetism formation}

\begin{figure}

\begin{tabular}{cc}
\includegraphics[scale=0.5]{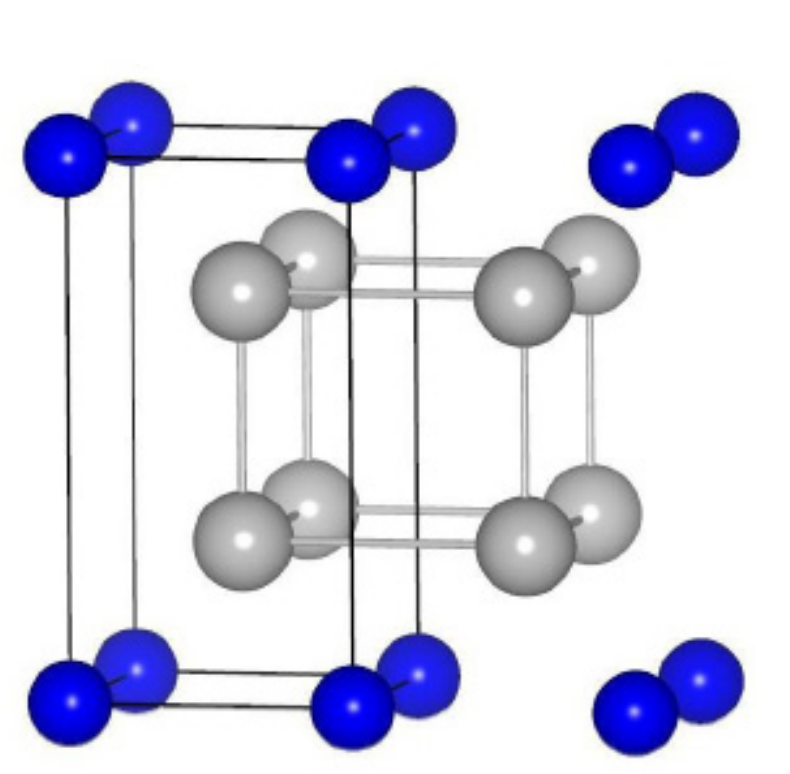} & \includegraphics[scale=0.55]{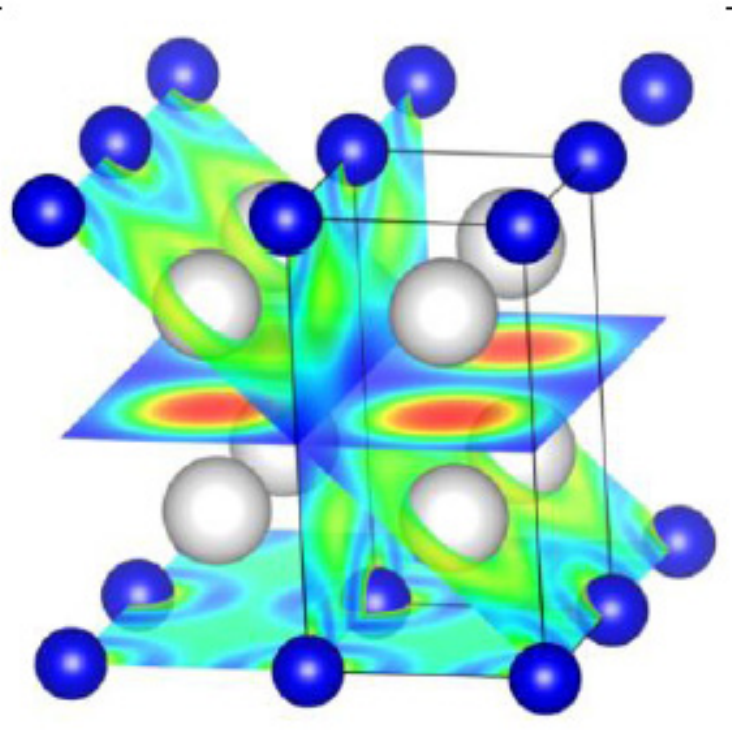}\tabularnewline
(a) & (b)\tabularnewline
\end{tabular}

\caption{Color online. (a) The crystal structure of $\alpha-\mathrm{FeSi}{}_{2}$;
(b) The electron localization function (ELF) for $\alpha-\mathrm{FeSi}{}_{2}$.
Blue and green colours correspond to the delocalized electrons, yellow
and red colours correspond to the localized electrons. The blue balls
stand for the Fe atoms, the grey ones are for the Si atoms \label{fig:Fig1}.}
\end{figure}

\begin{figure}
\begin{tabular}{c}
\includegraphics[scale=0.4]{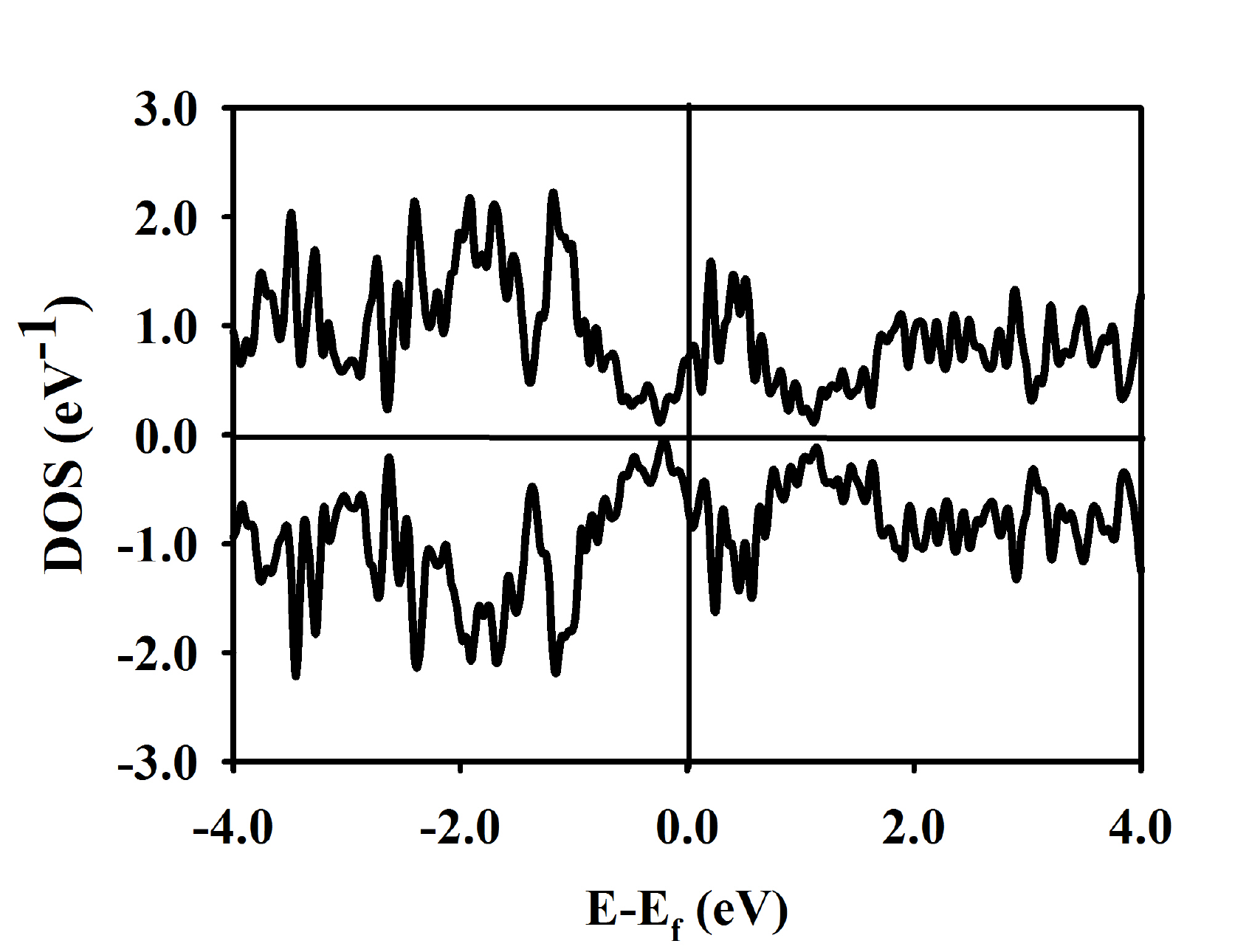}\tabularnewline
(a)\tabularnewline
\includegraphics[scale=0.4]{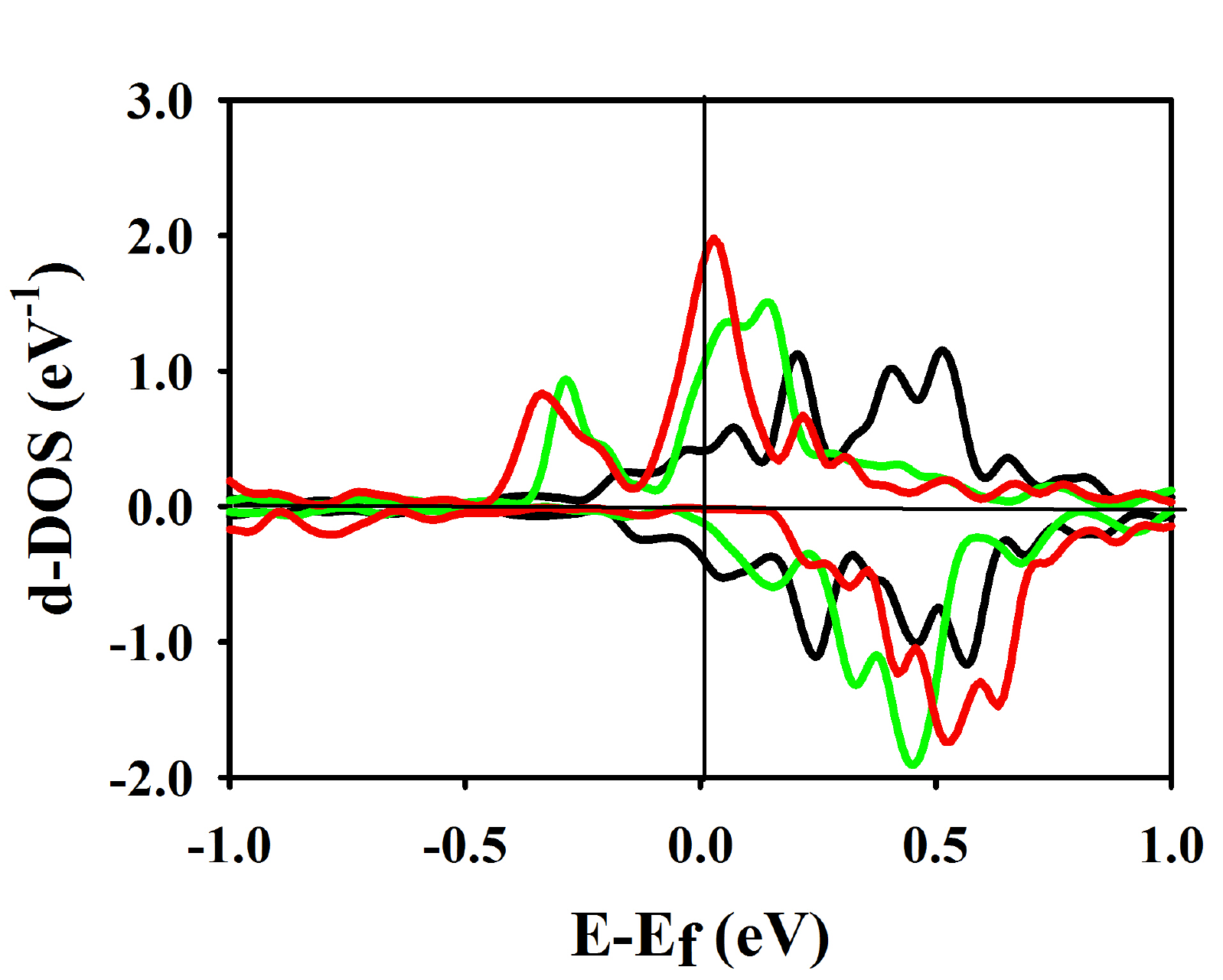}\tabularnewline
(b)\tabularnewline
\end{tabular}

\caption{Color online. (a) Full DOS for $\alpha-\mathrm{FeSi}{}_{2}$; (b) $e_{g}$-DOS
of $\alpha-\mathrm{FeSi}{}_{2}$ (black line), supercell with $R_{Fe-Fe}=2.73\,\mathring{A}$
(green line), supercell with $R_{Fe-Fe}=2.8\,\mathring{A}$ (red line).
Zero on the energy axis is the Fermi energy. \label{fig:Full-DOS}}

\end{figure}

The compound $\alpha-\mathrm{FeSi}{}_{2}$ has a tetragonal unit cell
with the lattice parameters $a=b=2.7\mathring{A}$ and $c=5.13\mathring{A}.$
Its structure is shown in Fig.\ref{fig:Fig1}a. The iron atoms are
located at $(0,0,0)$, the Si atoms are located at the points $(0.5,0.5,0,272)$
and $(0.5,0,5,0.728)$. As seen from Fig.\ref{fig:Fig1}a there is
a cavity between Si atoms in the structure due to the large distance
between Si atoms along the tetragonal $c-$axis ($R_{Si-Si}=2.4\mathring{A}$).
The calculated equilibrium distance between Fe -Si atoms  $R_{Fe-Si}=2.36\mathrm{\mathring{A}}$.
Our DFT-GGA calculations confirm that  the ground state of $\alpha-\mathrm{FeSi}{}_{2}$
is non-magnetic metal \cite{key-19}. The full density of electron
states (DOS) of $\alpha-\mathrm{FeSi}{}_{2}$ is shown in Fig. \ref{fig:Full-DOS}a.
The peak in the DOS in the vicinity of the Fermi energy is mainly
due to the $e_{g}$ $d-$electrons (fig.\ref{fig:Full-DOS}b, black
line). The other interesting peculiarity of the $\alpha-\mathrm{FeSi}{}_{2}$
structure is the presence of a network of quasi-one-dimensional channels,
which is easily seen on the map of electron localization function
\cite{key-20}(Fig.\ref{fig:Fig1}b).

While $\alpha-\mathrm{FeSi}{}_{2}$ in the bulk form is non-magnetic,
there are several experimental studies where ferromagnetism is found
in thin films \cite{key-7} and nano-particles \cite{key-8,key-9}.
Recently \cite{key-11} nano-sized grains $[001]$ - faceted $\alpha-\mathrm{FeSi}{}_{2}$
have been synthesized on a silicon substrate. The magnetic measurements
indicated the existence of small   magnetic moment (MM), $\sim0.2\mu_{B}$ per Fe atom.
According to experimental data \cite{key-11} the spacing between
Fe layers  along the tetragonal axis in the obtained nano-grains is
changed compared to that in the bulk: being larger between the layers
which are close to the substrate surface,  it decreases with distance
away from the substrate and then again increases. Simultaneously,
the stresses of $\sim1.2\%$ arise in the plane perpendicular to the
$c$ axis due to the misfit with the silicon substrate. These stresses
induce an increase of the distance $R_{Fe-Fe}$ between the iron atoms
in this plane.

In our earlier work \cite{key-10} we have shown that the ferromagnetism
can be induced by external stresses as well as by insertion into the
structure of additional atoms of iron or silicon. Contrary to the
suggestion \cite{key-8,key-9} which explains the appearance of the
magnetism in nanoparticles of $\alpha-\mathrm{FeSi}{}_{2}$ by formation
of Fe clusters, the results of our theoretical analysis, together
with experimental results \cite{key-11}, indicate that the stresses
alone may switch on the mechanisms  of the MM formation in $\alpha-\mathrm{FeSi}{}_{2}$.
In the present work we will investigate the influence of other types
of lattice distortions, on the ferromagnetism formation in $\alpha-\mathrm{FeSi}{}_{2}$.
Particularly, we expect that the MMs in $\alpha-\mathrm{FeSi}{}_{2}$
have to be sensitive to the changes of the distance between the layers
of irons and/or silicon.

In order to understand how the Fe interlayer distances $R_{IL}$ observed
in \cite{key-11} influence the magnetic properties of $\alpha-\mathrm{FeSi}{}_{2}$
we performed \emph{ab initio} calculations of the model supercell
$1\times1\times4$ with the different $R_{IL}$ (ranging from 5.13$\mathring{A}$
to 5.4$\mathring{A}$) between Fe planes along tetragonal axis. The
stress coming from the substrate are modeled by a $1.2\%$ increase
of the distance between in-plane iron atoms $(R_{Fe-Fe}=2.73\,\mathring{A}$)
compared to the one in the bulk $\alpha-\mathrm{FeSi}{}_{2}$ $(R_{Fe-Fe}=2.70\,\mathring{A})$.
The optimization of the supercell with respect to the atomic coordinates
results in the changing of the interlayer distances between Fe and
Si planes (on average by about 3\%) compared to the bulk ones. These
changes induce MMs about $0.2\mu_{B}$ on the Fe atoms in accordance
with experimental data. The partial contribution to DOS from the $e_{g}$-electrons
of Fe for this model supercell is shown by the green (on-line) curve
in Fig.\ref{fig:Full-DOS}b. The lattice distortion of the parent
$\alpha-\mathrm{FeSi}{}_{2}$ shifts the $e_{g}$-electron peaks in
the spin-up and spin-down DOS relative to each other and increases
the spin polarization by about $70\%$ in the model supercell. The
latter is one of the most important characteristics for the spintronic
applications. An increase of the lattice parameter up to $R_{Fe-Fe}=2.8$
$\mathring{A}$ leads to further amplification of these peaks in the
DOS and to the strong increase of the spin polarization (Fig.\ref{fig:Full-DOS}b,
red line). Thus, the \emph{ab intio} calculations indicate that increase
of merely the distance between in-plane Fe atoms results in the appearance
of small magnetic moments. In order to obtain moments of at least
$\simeq0.3\mu_{B}/atom$, the lattice parameter of $\alpha-\mathrm{FeSi}{}_{2}$
has to be increased by $~\sim5 \% $
($R_{Fe-Fe}=2.8\,\mathring{A}$ ), while an increase of the moment
till $\sim0.7\mu_{B}/atom$ requires the increase of the iron-iron
distance up to $\sim3\,\textrm{\ensuremath{\mathring{A}}}$, i.e.,
approximately, by $10 \% $ !
Although the distortions always arises when $\alpha-\mathrm{FeSi}{}_{2}$
film is experimentally synthesized on the Si\emph{ }substrate, it
never reaches such a large value. The experiment \cite{key-11}, however,
shows that the   MM $\sim0.2-0.3\,\mu_{B}/atom$ arises in the nanoparticles
of $\alpha-\mathrm{FeSi}{}_{2}$ at a smaller misfit strain, $\sim1.2\%$.
This fact prompts that, possibly, some other mechanisms of the moment
formation can be switched on by/during synthesis of the $\alpha-\mathrm{FeSi}{}_{2}$films.
The simplest ones are just other, different types of distortions.
There are several types of the bulk-$\alpha-\mathrm{FeSi}{}_{2}$-lattice
distortions which may cause the magnetism appearance in our model
supercells. It can be either an increase of the distance between Fe
atoms in the plane, or a change of the distance $R_{Si-Si}$ between
silicon atoms, or even the distance $R_{Fe-Si}$ between iron and
silicon atoms. Below we examine these possibilities in details.

\begin{figure}

\includegraphics[scale=0.4]{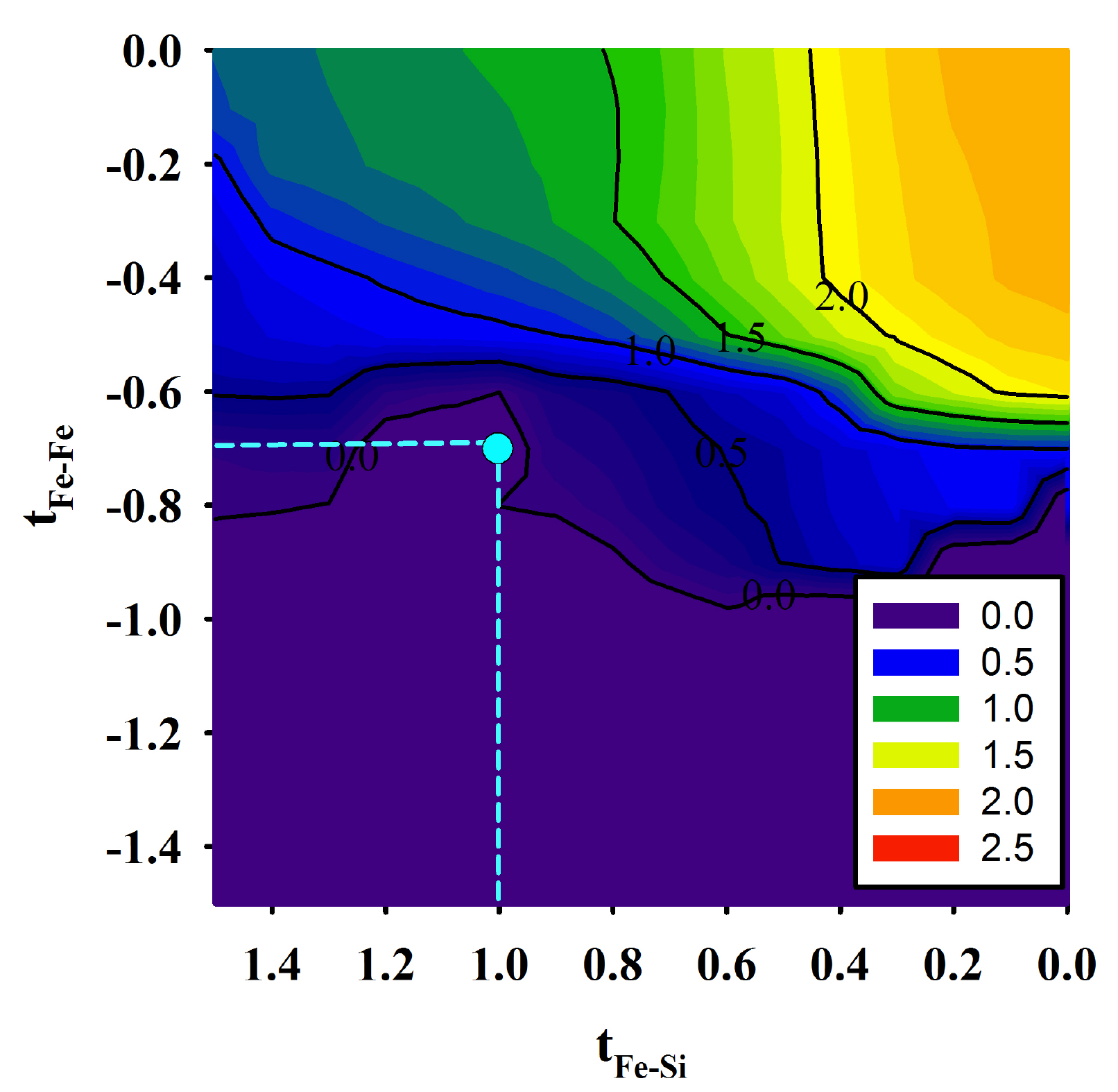}\caption{Color online. The map of magnetic moments $M(t_{Fe-Fe},t_{Fe-Si})$
for $\alpha-\mathrm{FeSi}{}_{2}$ at the equilibrium lattice parameter.
Dashed blue lines and blue point show the values of hopping integrals
which provide the best fit to the\emph{ ab initio} charge density.
The values of hopping parameters are given in eV.\label{fig:Map1}}

\end{figure}

A convenient tool for that is the mapping of the results, obtained
by the first-principle calculations, to the multiple-orbital model,
described in \cite{key-12} and shortly outlined in Sec. 2.2. According
to the results \cite{key-10}, the main parameter which controls the
MM formation is the hopping integral $t_{Fe-Fe}$ between the in-plane
Fe atoms ( Fig. \ref{fig:Map1}). Blue point on Fig.\ref{fig:Map1}
shows the values of hopping integrals ($t_{Fe-Fe}=-0.7 eV$, $t_{Fe-Si}=1.0 eV$,
$t_{Si-Si}=1.75 eV$) which provide the best fit to the\emph{ ab initio}
charge density for bulk $\alpha-\mathrm{FeSi}{}_{2}$.

\begin{figure}

\begin{tabular}{c}
\includegraphics[scale=0.6]{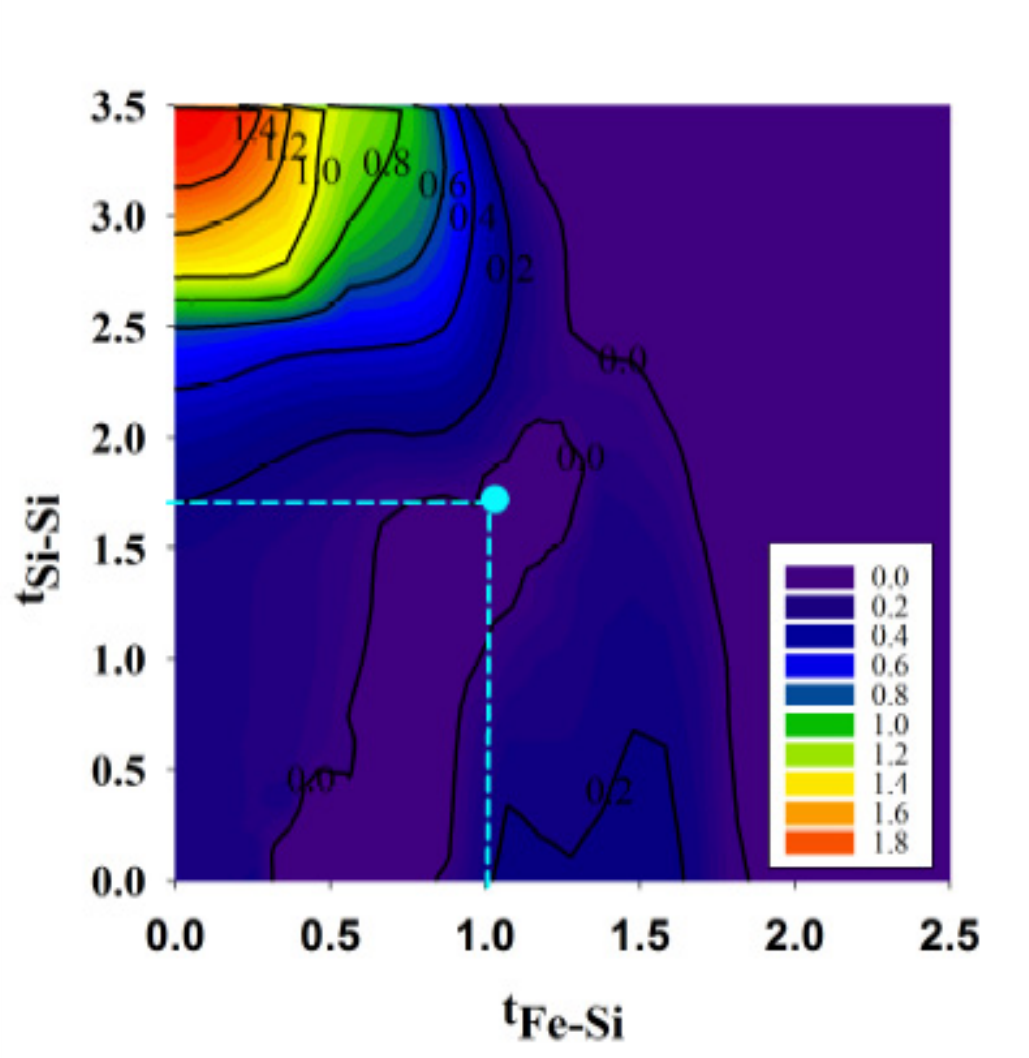}\tabularnewline
(a)\tabularnewline
\includegraphics[scale=0.6]{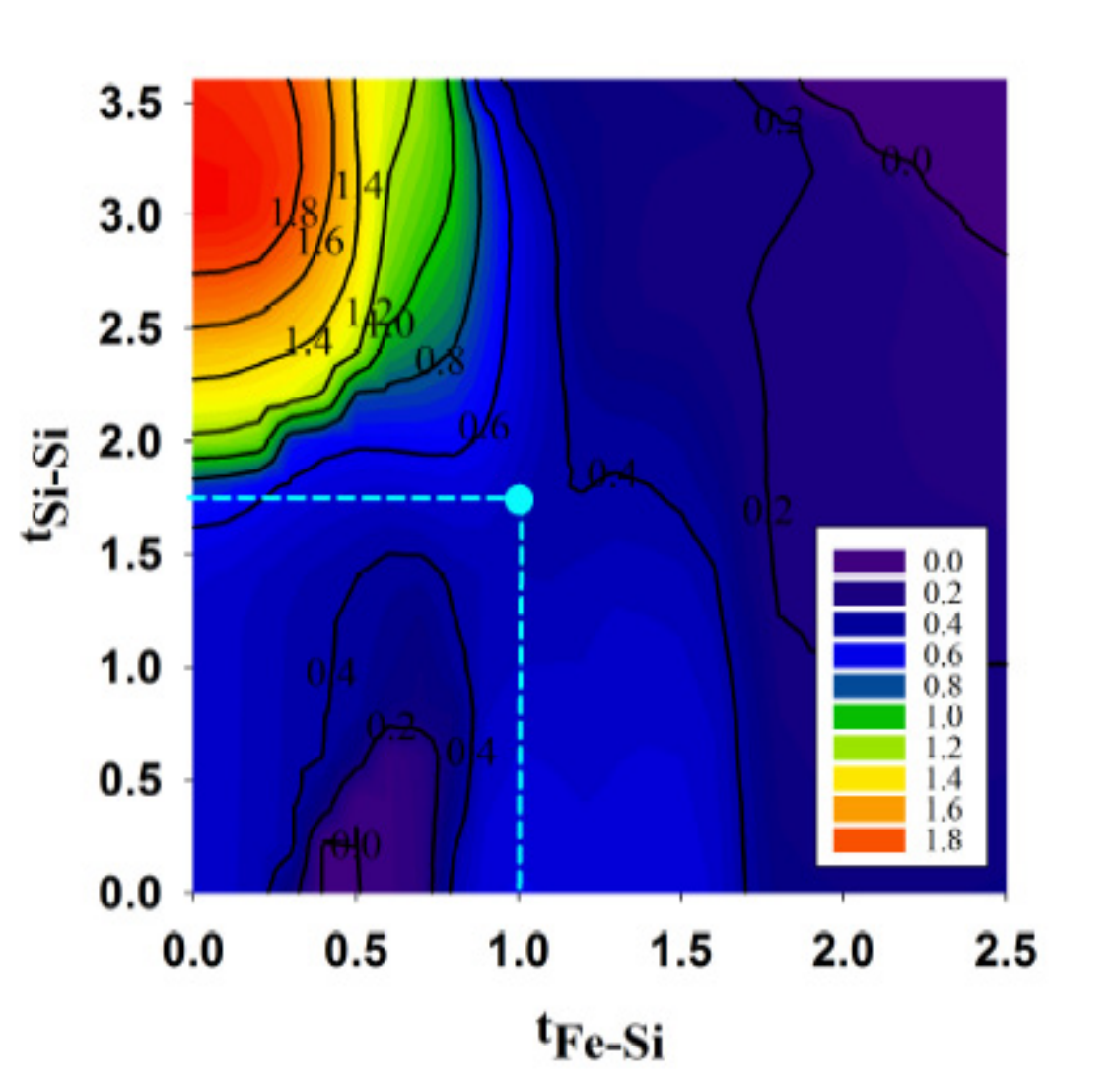}\tabularnewline
(b)\tabularnewline
\end{tabular}\caption{Color online. The map $M(t_{Fe-Si},t_{Si-Si})$ of   magnetic moments
$M$ for $\alpha-\mathrm{FeSi}{}_{2}$ : (a) at equilibrium lattice
parameter $a=2.7\mathring{A}$; (b) at $a=2.8\mathring{A}$. Dashed blue
lines and blue point show the values of hopping integrals which provide
the best fitting to the\emph{ ab initio} charge density. The values
of hopping parameters are given in eV.\label{fig:Map2}}

\end{figure}

The parameters for Fe - Si hopping $t_{Fe-Si}$ and Si - Si hopping
$t_{Si-Si}$ intuitively seem to be non-relevant to MM formation.
As will be seen below, this expectation is not supported by calculations.
Via the self-consistent solution of the model equations for the population
numbers of orbitals and the magnetization within the Hartree-Fock
approximation we obtained the MM map in the coordinates $t_{Fe-Si}$
vs $t_{Si-Si}$ at the fixed value of $t_{Fe-Fe}=-0.7 eV$ (Fig.\ref{fig:Map2}a).
The latter value corresponds to the equilibrium Fe - Fe distance $R_{Fe-Fe}=2.7\textrm{\ensuremath{\mathring{A}}}$
for bulk $\alpha-\mathrm{FeSi}{}_{2}$. Notice that a decrease of
the distance between silicon atoms, $R_{Si-Si}$, increases the distance
between Fe and Si atoms, and \emph{vice versa }(Fig.1a)\emph{.} As
seen from the map in Fig.\ref{fig:Map2}a, there is no magnetism at
the equilibrium distance $R_{Fe-Fe}=2.7\textrm{\ensuremath{\mathring{A}}}$
in undistorted $\alpha-\mathrm{FeSi}{}_{2}$. However, a decrease
of the hopping integral $t_{Fe-Si}$ with simultaneous increase of
$t_{Si-Si}$ leads to the arising of the magnetism \emph{at the same
distance} $R_{Fe-Fe}$. As seen from the upper left corner of the
map Fig.(\ref{fig:Map2}a), a large   MM $\sim1-1.1\mu_{B}$ can be
achieved by decrease of the distance between Si atoms which causes
the changes of the hopping integral magnitudes. So, hoppings integrals
$t_{Si-Si}\approx3.1$$\:$ and $t_{Fe-Si}\approx0.5$ correspond
to distances $R_{Si-Si}\approx1.6\mathring{A}$ and $R_{Fe-Si}\approx2.6\mathring{A}$.
And \textsl{vice versa} an increase of the $Si-Si$ distance (decrease
of $t_{Si-Si}$ and increase of $t_{Fe-Si}$ ) leading to decrease
of the MM to $\sim0.1-0.3\mu_{B}$.

\begin{figure}
\includegraphics[scale=0.45]{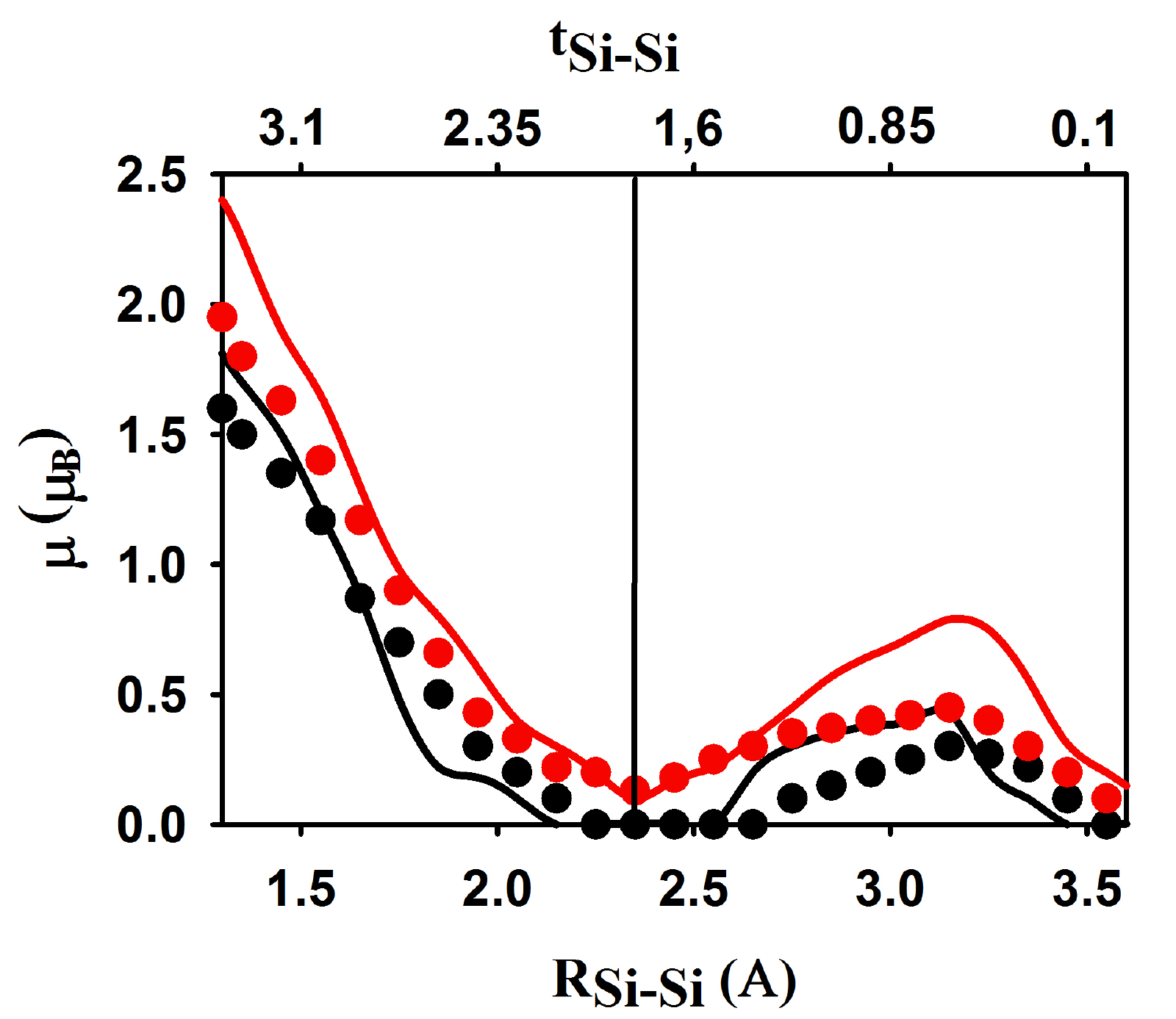}

\caption{Color online. The dependence of the   MM in $\alpha-\mathrm{FeSi}{}_{2}$on
the distance $R_{Si-Si}$ between silicon atoms ( the hopping integrals
in the model $t=t(R_{Si-Si})$ ). The results for $R_{Fe-Fe}=2.7\mathring{A}$
are displayed by the black color and for $R_{Fe-Fe}=2.78\mathring{A}$
by the red one. The solid line stands for GGA (in VASP), the points
are for the model within the HFA. The vertical line indicates the
equilibrium distance $R_{Si-Si}=2.34\mathring{A}$ in $\alpha-\mathrm{FeSi}{}_{2}$.
The values of hopping parameters are given in eV.\label{fig:M(R_SiSi)}}
\end{figure}

Thus, the analysis of the model within the Hartree-Fock approximation
shows that the ferromagnetic state in $\alpha-\mathrm{FeSi}{}_{2}$
may be induced by: (a) the increase of the distance between iron atoms
(Fig.3), and (b) the change of the distance between NN silicon and
iron atoms and between silicon atoms (Fig.4a). An application of both
types of changes expands the area of existence of the ferromagnetic
solutions. This is illustrated by Fig.\ref{fig:Map2}b, which displays
the map of   MMs evaluated at $t_{Fe-Fe}=-0.65 eV$ . This corresponds
to $R_{Fe-Fe}=2.78\textrm{\ensuremath{\mathring{A}}}$, according
to Eq.(5), i.e., to the misfit strain $\sim3\%.$ At this distance the
magnitudes of the   MM $M\sim1.0\mu_{B}$ arise at the smaller hoppings
(Fig.4b) $t_{Si-Si}\approx2.7$ ($R_{Si-Si}\approx1.8\mathring{A}$)
and\textbf{ }$t_{Fe-Si}\approx0.65$ ($R_{Fe-Si}\approx2.5\mathring{A}$\textbf{).}
In order to confirm the model findings we performed \emph{ab initio}
calculations of the moment dependence on the distances between silicon
atoms in $\alpha-\mathrm{FeSi}{}_{2}$.  Fig.\ref{fig:M(R_SiSi)} displays
the comparison of the results of model and \emph{ab initio} calculations
for the dependence of the MM at iron atoms on silicon-silicon distance
$R_{Si-Si}$ at equilibrium and expanded $R_{Fe-Fe}$ distances between
in-plane iron atoms. Similar to the model result, a decrease of $R_{Si-Si}$
causes a sharp increase of the MM. Notice that a slight increase of
$R_{Si-Si}$ also may induce MM, but in this case the moment is small.
An increase of the distance between iron atoms leads to appearance
of a large moment at the same distance $R_{Si-Si}$. Fig. \ref{fig:M(R_SiSi)}
confirms that the results of the model and the \emph{ab initio} calculations
are in quite good agreement with each other.

\begin{figure}

\includegraphics[scale=0.45]{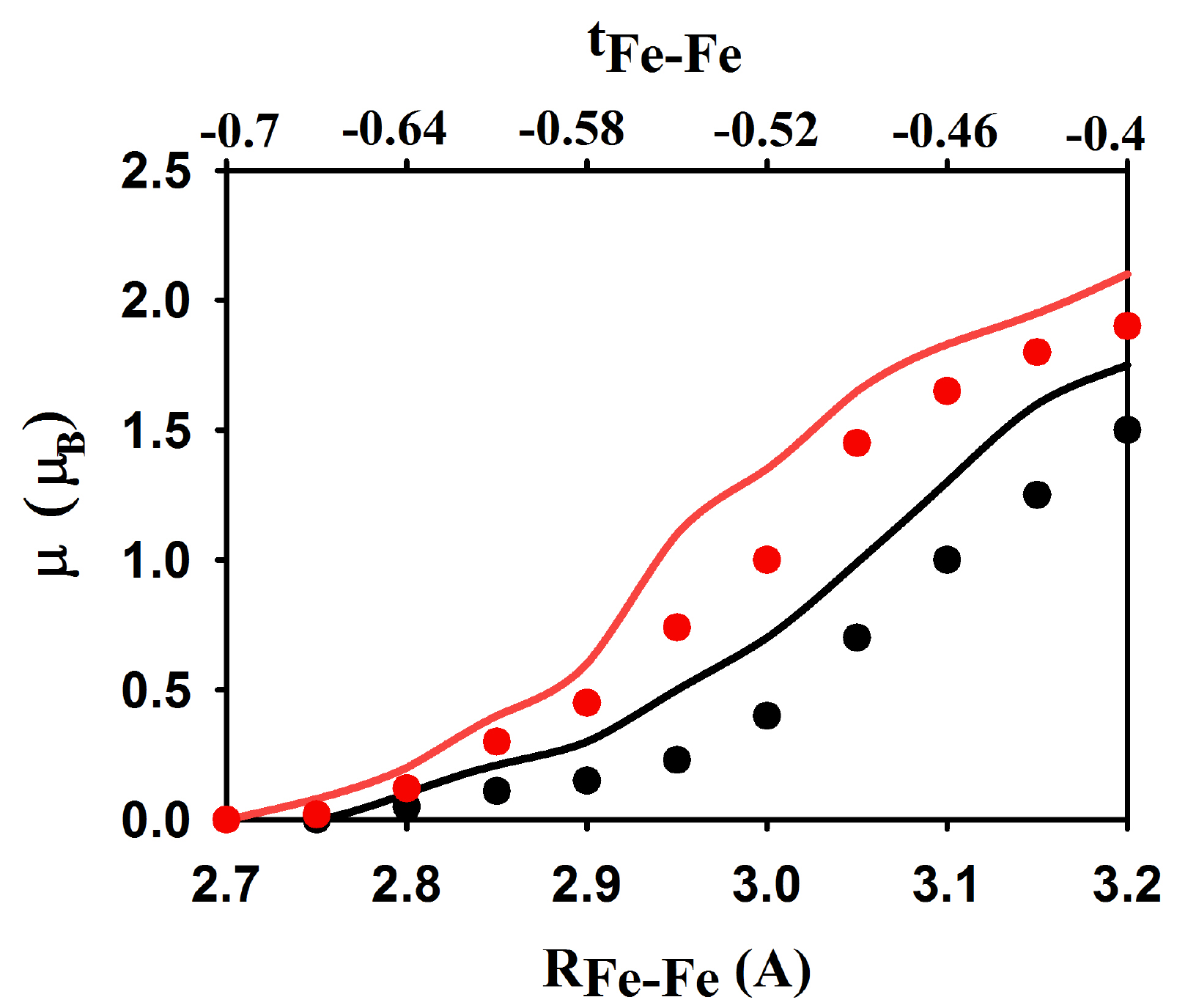}\caption{Color online. The dependence of the MM in $\alpha-\mathrm{FeSi}{}_{2}$
 on the distance $R_{Fe-Fe}$ between iron atoms ( the hopping integrals
in the model $t=t(R_{Fe-Fe})$ ). The results for $R_{Si-Si}=2.34\textrm{?}$
are displayed by the black color and for $R_{Si-Si}=1.90\textrm{\ensuremath{\mathring{A}}}$
by the red one. The solid line stands for GGA (in VASP), the points
are for the model within the HFA. \label{fig:MofDifferentR}}

\end{figure}

The analysis performed in this part can be summarized as follows.
Both the model and \emph{ab initio} calculations indicate that the
ferromagnetism in $\alpha-\mathrm{FeSi}{}_{2}$ can be induced by different
types of the lattice distortions: not only by an increase of the in-plane
distance between iron atoms, but also by a change of the distance
between layers along the tetragonal axis. The latter alters iron-silicon
and silicon-silicon interatomic distances (see Fig.\ref{fig:M(R_SiSi)}
). The decisive parameter for MM formation is the iron-iron distance
in the plane perpendicular to the tetragonal axis \emph{c} (Fig.\ref{fig:Map1},
\ref{fig:Map2}). However, in order to obtain the moments large enough
for practical applications, the required misfit strain has to be made
quite large, $\sim10-15\%.$ Such big magnitudes can hardly be achieved
experimentally. At the experimentally feasible range of the misfit
strain $~1-3\%$ the MM remains to be small. The other solution would
consists of simultaneous decrease of the Si - Si distance ($R_{Si-Si}$)
and increase of the Fe - Si and Fe - Fe distances. Indeed, as seen
from Fig.\ref{fig:MofDifferentR}, where the dependence of the on-iron-MM
on the distance $R_{Fe-Fe}$ at $R_{Si-Si}=2.34\mathring{A}$
and $R_{Si-Si}=1.9\mathring{A}$ is displayed; the decrease of $R_{Si-Si}$
gives rise to a larger MM at the same Fe - Fe distance.

\begin{figure}

\includegraphics[scale=0.5]{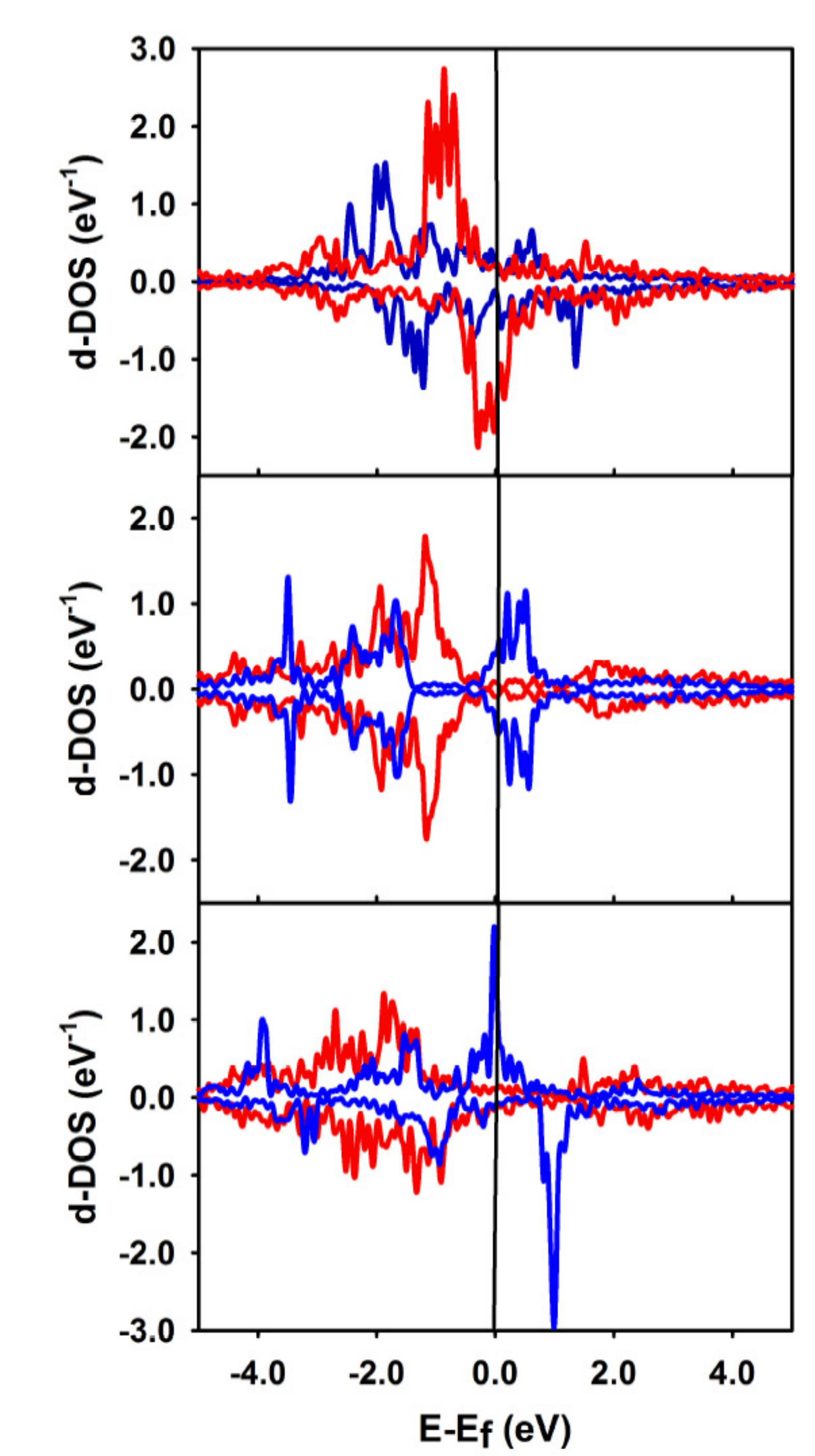}\caption{Color online. Partial densities of $d$-states of $\alpha-\mathrm{FeSi}{}_{2}$
for different lattice distortions. From top to bottom:  $R_{Si-Si}=1.90$$\mathring{A}$,
the equilibrium $R_{Si-Si}=2.34$$\mathring{A}, $ $R_{Si-Si}=2.90$$\mathring{A}$.
The $t_{2g}$-DOS is displayed by red (on-line) and the $e_{g}$ ones
by blue colors. Zero on the energy axis is the Fermi energy. \label{fig:PartialDOS} }

\end{figure}

Another way to understand why some of the lattice distortions favor
to the magnetism appearance is to analyze the evolution of the partial
density of $d$-electron states ($d$-DOS) with these distortions.
As shown at Fig.\ref{fig:PartialDOS} the decrease of the distance
between silicones ($R_{Si-Si}$) shifts the peaks of the $t_{2g}^{\downarrow}$
states, which move towards the Fermi level. This, in turn, gives rise
to spin polarization. However, an increase of $R_{Si-Si}$ or the
distance $R_{Fe-Fe}$ between the iron atoms shifts not the $t_{2g}$-,
but the $e_{g}$-peaks. In this case the $e_{g}^{\uparrow}$-states
appear near the Fermi level, providing a non-zero spin polarization.

The analysis given above highlights the main difficulty which is expected
to arise in experiment on inducing a magnetism in  $\alpha-\mathrm{FeSi}{}_{2}$
with reasonably large MM via the lattice distortions. Particularly,
the MM $\simeq1.0\mu_{B}$ should arise at $R_{Fe-Fe}\simeq3\mathring{A}$,
or $R_{Si-Si}\simeq1.8\mathring{A}$ ($R_{Fe-Si}\simeq2.5\textrm{\ensuremath{\mathring{A}}}$
). Such distances between atoms are hardly possible to implement in
$\alpha-\mathrm{FeSi}{}_{2}$ films with any type of substrate. The
distortions which arise when the $\alpha-\mathrm{FeSi}{}_{2}$ is
grown on the silicon substrate are much smaller: in the experiments
\cite{key-11} on $\alpha-\mathrm{FeSi}{}_{2}$ nanoparticles the
magnitudes of the distortions between in-plane iron atoms are about
$1\%$, while for interlayer distances this is about $5\%$. Such
small distortions induce, correspondingly, small MM. The question
arises, would it be possible to overcome this difficulty with a ``chemical
pressure''?

\subsection{The effect of intercalation on the magnetism formation}

\begin{widetext}

\begin{figure}

\begin{tabular}{cc}
\includegraphics[scale=0.75]{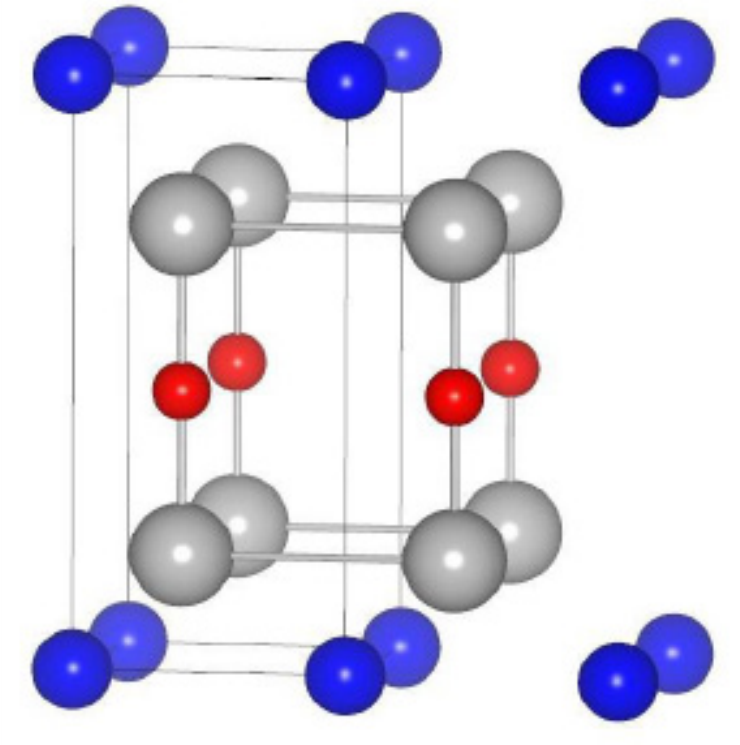} & \includegraphics[scale=0.7]{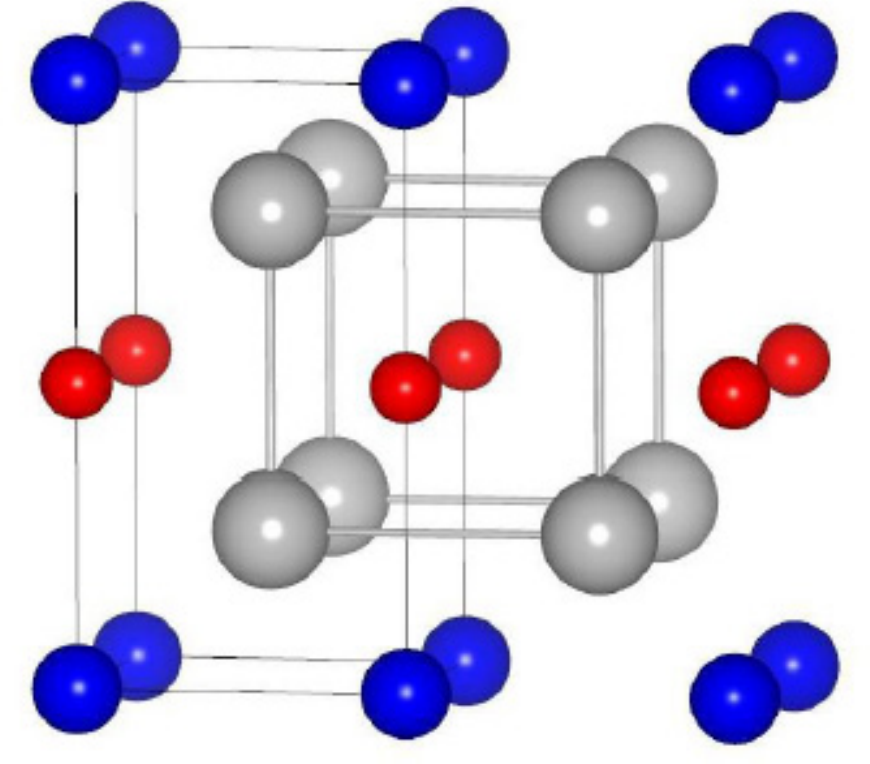}\tabularnewline
a & b\tabularnewline
\end{tabular}\caption{Color online. Two possible positions for atom imbedding into the cavity
between silicon atoms. (a) The positions occupied by non-metal atoms;
(b) The positions occupied by metal atoms. Blue balls represent the
Fe atoms, grey balls represent Si atoms, red balls represent intercalated
atoms.\label{fig:EmbAtPositions}}
\end{figure}

\end{widetext}

As was mentioned above, there is a cavity between Si atoms in the
$\alpha-\mathrm{FeSi}{}_{2}$ structure (Fig.1a). An intercalation
of other atoms into this cavity will distort the lattice. Here we
investigate if intercalated atoms can introduce the change of the
distances $R_{Fe-Fe}$ and $R_{Si-Si}$ sufficient for magnetism appearance.
In order to check this hypothesis we performed \emph{ab initio} calculations
of $\alpha-\mathrm{FeSi}{}_{2}$ with embedded atoms of different
elements. The results suggest that there are two positions for guest
atoms which are the most energetically favorable. The types of positions
for embedding the guest atoms are shown at Fig.\ref{fig:EmbAtPositions}.
Notice that all considered structures have been fully optimized.

The non-metal atoms are found to prefer the positions on the bonds
between the silicon atoms (Fig.8a), whereas the position inside of
the tetragonal cavity formed by the silicon atoms is more energetically
favorable for the atoms of metals (Fig.\ref{fig:EmbAtPositions}b).
The intercalated atoms create a negative chemical pressure which results
in an increase of the distance between host atoms compared to pure
$\alpha-\mathrm{FeSi}{}_{2}$.  The results of calculations are summarized
in the Table \ref{tab:Tab1}, where the parameters of the lattice
cell, the values of the MMs at the iron atoms and the spin polarization
in some of the considered structures are shown.

\begin{widetext}
\begin{center}
\begin{table}
\caption{The lattice parameters ($a,c$), magnetic moments on Fe atom ($\mu_{Fe}$),
the distance between Fe - Si ($R_{Fe-Si}$) and Si - X (X - intercalant)
atoms ($R_{Si-X}$), spin polarization ($P=\frac{\rho^\uparrow(\varepsilon_F)-\rho^\downarrow(\varepsilon_F)}{\rho^\uparrow(\varepsilon_F)+\rho^\downarrow(\varepsilon_F)}\cdot100\%$) and anisotropy of the
plasma frequency $\eta=\Omega_{xx}/\Omega_{zz}$ in the intercalated
$\alpha-\mathrm{FeSi}{}_{2}$\label{tab:Tab1} }

\begin{tabular}{|c|c|c|c|c|c|c|c|}
\hline
\multirow{1}{*}{Atom X} & \multicolumn{2}{c|}{Lattice parameters ($\mathring{A}$)} & \multirow{1}{*}{$\mu_{Fe}(\mu_{B})$} & \multirow{1}{*}{$R_{Fe-Si}(\mathring{A})$} & \multirow{1}{*}{$R_{Si-X}(\mathring{A})$} & \multirow{1}{*}{$P( \% ) $} & $\eta=\Omega_{xx}/\Omega_{zz}$\tabularnewline
\hline
\multicolumn{8}{|c|}{\textbf{Position 1} (Fig.\ref{fig:EmbAtPositions}\textcyr{\char224})}\tabularnewline
\hline
H & 2.72  & 6.28 & 0.20 & 2.37 & 1.76 & 0 & 0.52\tabularnewline
\hline
O & 2.76 & 5.91 & 0.45 & 2.36 & 1.63 & 61 & 1.76\tabularnewline
\hline
P & 2.71 & 7.27 & 0.47 & 2.28 & 2.35 & 75 & 1.63\tabularnewline
\hline
As & 2.76 & 7.58 & 0.53 & 2.36 & 2.46 & 38 & 1.73\tabularnewline
\hline
Sb & 2.89 & 7.80 & 0.90 & 2.40 & 2.64 & 38 & 2.12\tabularnewline
\hline
N & 2.74 & 6.05 & 0.00 & 2.39 & 1.63 & 0 & 1.96\tabularnewline
\hline
\multicolumn{8}{|c|}{\textbf{Position 2} (Fig.\ref{fig:EmbAtPositions}b)}\tabularnewline
\hline
Li & 2.93 & 5.34 & 0.90 & 2.43 & 2.50 & 11 & 0.84\tabularnewline
\hline
Na & 2.90 & 6.80 & 0.65 & 2.41 & 2.94 & 71 & 1.19\tabularnewline
\hline
K & 2.87 & 8.07 & 0.14 & 2.44 & 3.36 & 68 & 2.68\tabularnewline
\hline
Ca & 3.00 & 6.85 & 1.00 & 2.46 & 3.04 & 11 & 1.03\tabularnewline
\hline
Sr & 3.09 & 7.02 & 1.05 & 2.50 & 3.16 & 11 & 1.09\tabularnewline
\hline
Cu & 2.95 & 5.38 & 0.90 & 2.24 & 2.52 & 24 & 1.06\tabularnewline
\hline
$\alpha-\mathrm{FeSi}{}_{2}$ & 2.70 & 5.13 & 0.00 & 2.36 & - & 0 & 0.87\tabularnewline
\hline
\end{tabular}
\end{table}
\end{center}
\end{widetext}

As seen from Table I, an intercalation not always leads to magnetic
state formation. \emph{E.g}., the structures with the intercalated
nitrogen atoms are not magnetic (see Tab.\ref{tab:Tab1}). Nevertheless,
the general tendency of the MM increase with the increase of lattice
distortions, studied in the previous Section, is reproduced by the
direct calculation. As expected, the magnitude of the MM at iron atoms
grows with an increase of the Fe - Fe distance, but an increase of
$R_{Fe-Fe}$ only does not provide the magnitudes of MM listed in
the Tab.\ref{tab:Tab1}. For example, an increase of in-plane distance
$R_{Fe-Fe}$ up to $2.95\mathring{A}$ in pure $\alpha-\mathrm{FeSi}{}_{2}$
leads to MM at Fe atom $M_{Fe}\simeq0.5\mu_{B}$ only (Fig.\ref{fig:MofDifferentR})
whereas intercalation by some atoms increases MM $1.5-2$ times at
the same\textbf{ $R_{Fe-Fe}$.} The latter occurs due to additional
structure distortions and corresponding restructuring of DOS due to
intercalated atoms.

\begin{widetext}

\begin{figure}

\begin{tabular}{cc}
\includegraphics[scale=0.5]{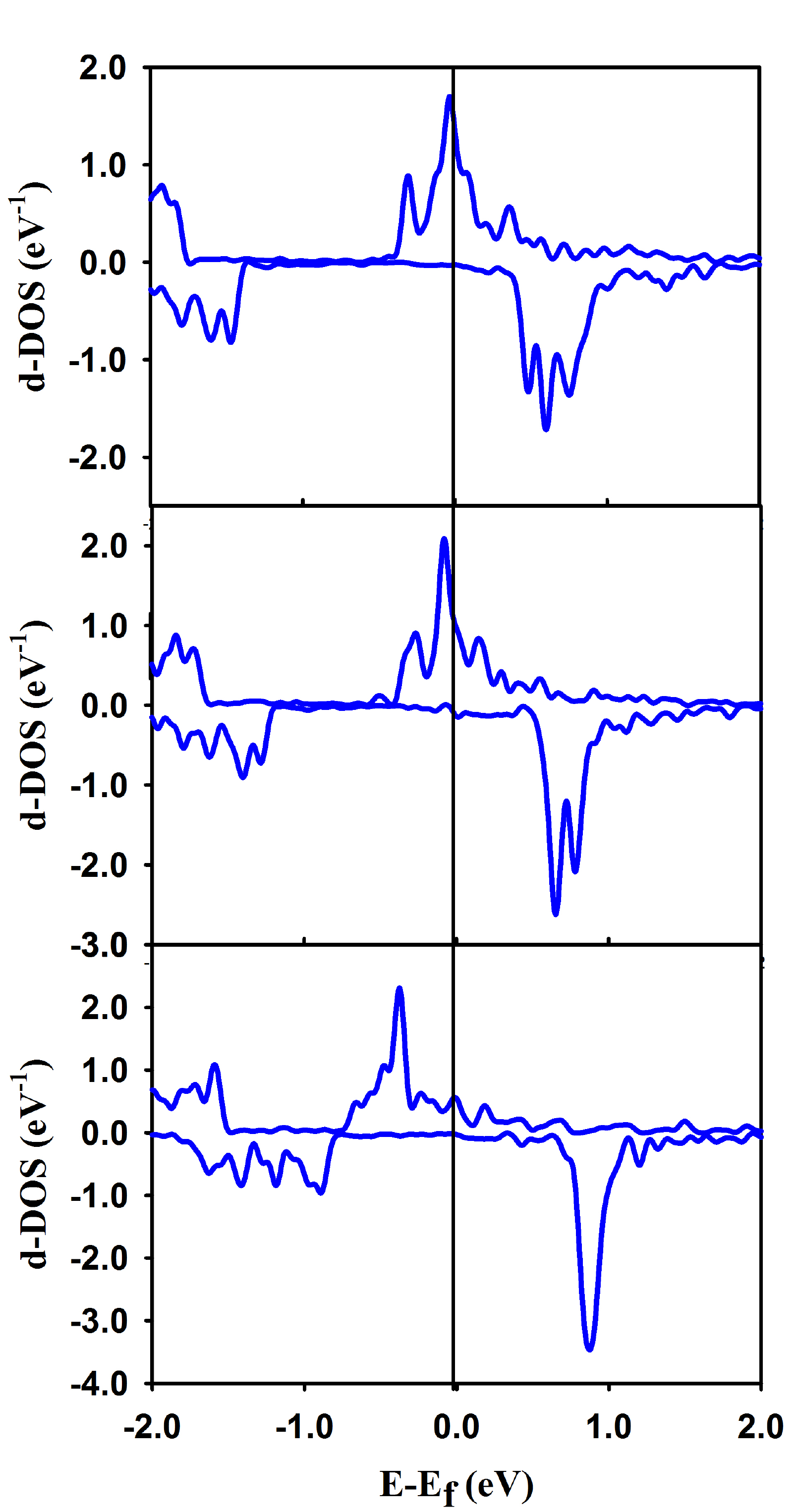} & \includegraphics[scale=0.5]{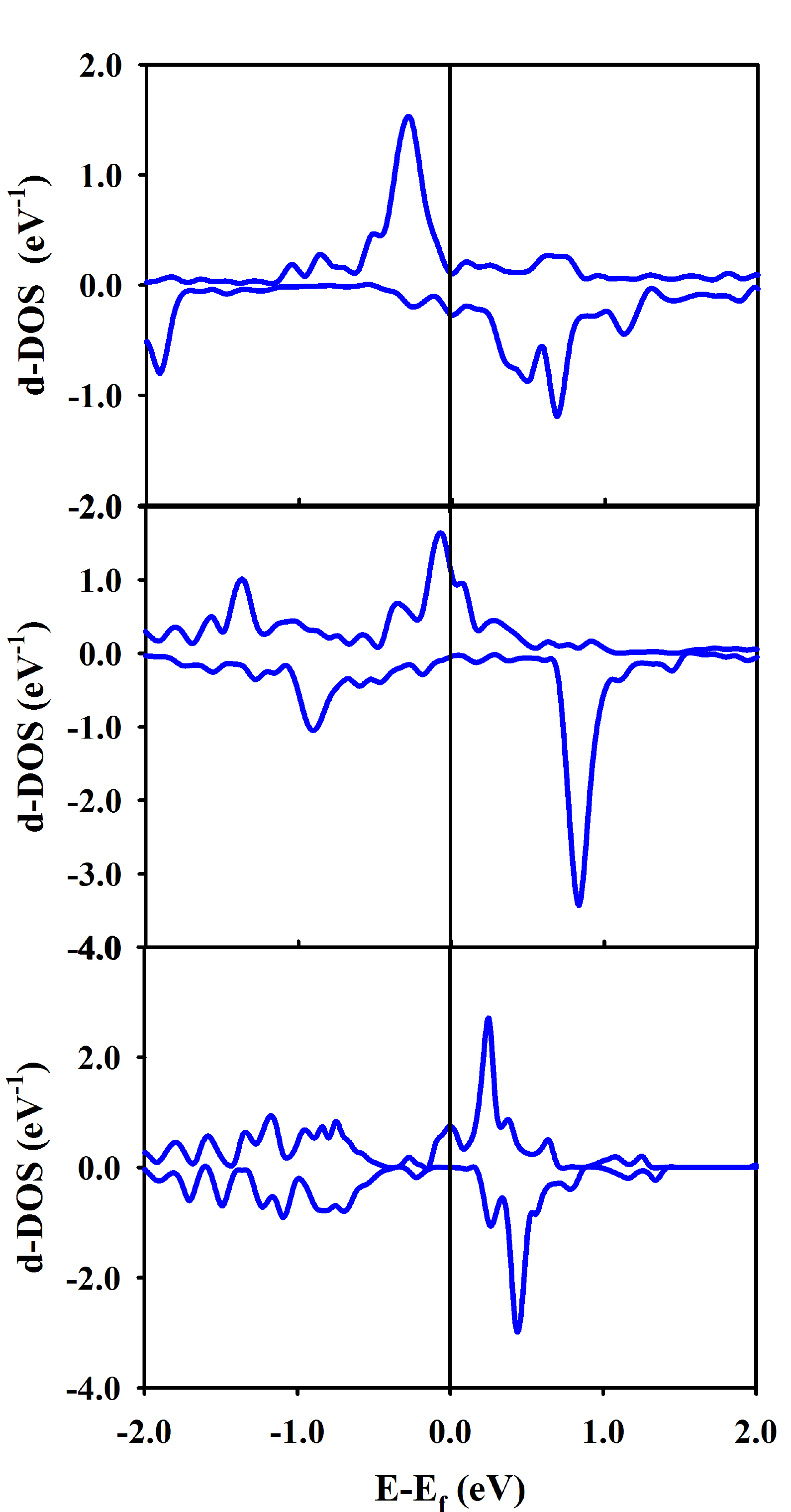}\tabularnewline
(a) & (b)\tabularnewline
\end{tabular}\caption{(a) $e_{g}$-DOS for $\alpha-\mathrm{FeSi}{}_{2}$ intercalated by
the non-metal atoms which occupy the first positions in Fig.  \ref{fig:EmbAtPositions};
from top to bottom: P, As, Sb (b) $e_{g}$-DOS for $\alpha-\mathrm{FeSi}{}_{2}$
intercalated by the metal atoms (second position in Fig. \ref{fig:EmbAtPositions});
from top to bottom: Li, Na, K Zero on the energy axis is at the Fermi
energy. \label{fig:DOSwIntercalation}}

\end{figure}

\end{widetext}

The calculations show that the distortions caused by intercalation
change the $t_{2g}$-DOS only slightly, the main changes occur in
the $e_{g}$-DOS. Similar to pure $\alpha-\mathrm{FeSi}{}_{2}$ it
is namely the $e_{g}$-DOS that forms the peaks in the vicinity of
the Fermi level. This is illustrated in Fig.\ref{fig:DOSwIntercalation}
for several intercalates: similar to pure $\alpha-\mathrm{FeSi}{}_{2}$
the increase of the distances $R_{Fe-Fe}$ and $R_{Si-Si}$ causes
shifts of the $e_{g}^{\uparrow}$- and $e_{g}^{\downarrow}$- peaks.
In the case on non-metallic intercalates (P, As, Sb) this shift grows
with an increase of the distances $R_{Fe-Fe}$ and $R_{Si-X}$(Fig.\ref{fig:DOSwIntercalation}a).
For the metallic intercalates Li, Na, K the tendency is opposite (Fig.\ref{fig:DOSwIntercalation}b).
As seen from Fig.\ref{fig:DOSwIntercalation} and Table \ref{tab:Tab1},
the intercalation by $P$ and $Na$ is expected to provide high spin
polarization due to crossing the Fermi energy by the $d^{\uparrow}$-
peaks of DOS. The quite strong spin polarization (7-th column in Table
\ref{tab:Tab1}) may occur not only in the cases of intercalation
by mentioned above P and Na , but also by O and K. Small increase
of the in-plane lattice parameter which arisen when the $\alpha-\mathrm{FeSi}{}_{2}$
is intercalated by H, As, O or P allows for use of the silicon substrate.
The intercalation by Li, Na, K and Sb atoms results in the 7\% increase
of the in-plane lattice parameter compared to pure $\alpha-\mathrm{FeSi}{}_{2}$,
but the compressive strain from the substrate can decreases this distortion.
This, however, does not kill MM completely, but decreases it by 30-40\%.
Therefore, one can expect that the choice of a substrate with a larger
lattice parameter than that of silicon (e.g., Ge) would allow to decrease
this misfit strain and increase the magnitude of MM.

Since it is hardly possible to achieve 100\% concentration of intercalated
atoms in the experiment, we estimate the value of MM arising at the
Fe atoms for the lower concentration of intercalated atoms, namely
for $25$\% and $50$\% concentrations of intercalated atoms. In order
to consider a possible ordering of intercalated atoms with these concentrations\textbf{
}we constructed a\textbf{ $2\times2\times2$ }supercell of $\alpha-\mathrm{FeSi}{}_{2}$\textbf{.
}Further calculations depend on the way how the sample is made. An
annealing of a sample may switch on the thermodynamic equilibration
processes, possibly, ion migration, \emph{etc}.. This may exclude
the contribution of the less energetically favorable configurations.
An estimation of the barriers for migration of ions and the contribution
of phonons is needed for a quantitative description of these processes.
This requires special consideration. In the case when a sample is
made by quenching the situation is simpler: due to fast cooling of
the sample the energy hierarchy of different possible configurations
is much less important and their contributions to an averaged physical
quantity $<A>$ may be calculated either with the help of simple statistical
weights, or by means of some of realization of the coherent potential
approximation. The latter, however, also involves additional assumption
about the distribution function and the way, how the effective medium
is introduced, but has the advantage that it does not require supercell
calculations and can be used for arbitrary (but not too small) concentration.
Here we consider the first case.

Let us denote the statistical weights of the configurations $i$ as
$w_{x}^{(n)}(i)$, where $x$ is the concentration of intercalated
atoms, and $n_{x}^{i}$ is number of equivalent configurations of
the type $i$ , and $A_{x}^{n}(i)$ is value of the physical quantity
in the configurations $i$. Then the total number of configurations
is $N_{c}(x)=\sum_{i}n_{x}^{i}w_{x}^{(n)}(i)$ and
\begin{equation}
<A>=\frac{1}{N_{c}(x)}\sum_{i}A_{x}^{n_{i}}(i)n_{x}^{i}w_{x}^{(n)}(i).\label{eq:AvOverConf}
\end{equation}

For the $25\%$ concentration of intercalating atoms we have $N_{c}(0.25)=\left(\begin{array}{c}
8\\
2
\end{array}\right)=28$ arrangements. Five of them are different. Considering all possible
configurations we find that for the $x=0.25$ the statistical weight
$w_{0.25}^{(3)}=4$ for tree of them, and $w_{0.25}^{(2)}=8$ for
two of them. For $50\%$ concentration of intercalated atoms there
are $N_{c}(0.5)=\left(\begin{array}{c}
8\\
4
\end{array}\right)=70$ possible ways to distribute atoms. Nine of them are different: for
three configurations $w_{0.5}^{(3)}=2$, for two it is $w_{0.5}^{(2)}=4$,
$w_{0.5}^{(2)}=8$,  $w_{0.5}^{(1)}=16$, and $w_{0.5}^{(1)}=24$
($\sum_{i}n_{0.5}^{i}w_{0.5}^{(n)}=70$). We find that the perspective
candidates are Li metal and the O non-metal intercalates. As seen
from Table \ref{tab:Tab1}, the intercalation by these atoms results
in the relatively large MMs for comparatively small lattice distortions.
We performed the full optimization for all ordered structures. In
Tab.\ref{tab:Tab2} we give the difference $\Delta E=E_{max}-E_{min}$
between maximal and minimal energies of the structures for each of
cases. The total energy values of different ordered structures are
within $0.5$ eV range per unit cell of $\alpha-\mathrm{FeSi}{}_{2}$.
The average lattice parameters $<a>,<c>(\mathring{A})$, and Fe MMs
$<\mu>(\mu_{B})$ and spin polarization$<P>(\%)
$, calculated according to Eq. \ref{eq:AvOverConf} are given in Table
\ref{tab:Tab2}.

\begin{table}
\caption{The energy difference ($\varDelta E$) between maximal and minimal
energies of ordered supercells of intercalated $\alpha-\mathrm{FeSi}{}_{2}$,
average magnetic moments ($<\mu>$), spin polarization ($<P>$) and
lattice parameters ($<a>,$ $<c>$) in ordered intercalated $\alpha-\mathrm{FeSi}{}_{2}$
for 25\% and 50\% concentrations of intercalated atoms of oxygen and
lithium.\label{tab:Tab2}}
\begin{tabular}{|c|c|c|c|c|}
\hline
 & \multicolumn{2}{c|}{O} & \multicolumn{2}{c|}{Li}\tabularnewline
\hline
\hline
 & 25\% & 50\% & 25\% & 50\%\tabularnewline
\hline
$\varDelta E$ (eV) & 0.23 & 0.5 & 0.05 & 0.42\tabularnewline
\hline
$<\mu>(\mu_{B})$ & 0.14 & 0.28 & 0.31 & 0.67\tabularnewline
\hline
$<P>(\%)
$ & 61 & 63 & 57 & 35\tabularnewline
\hline
$<a>(\mathring{A})$ & 2.72 & 2.74 & 2.77 & 2.82\tabularnewline
\hline
$<c>(\mathring{A})$ & 5.43 & 5.66 & 5.14 & 5.19\tabularnewline
\hline
\end{tabular}
\end{table}

Although at $25$\% concentration of lithium atoms the lattice parameters
have only little change as compared with $\alpha-\mathrm{FeSi}{}_{2}$
(Tab.\ref{tab:Tab2}), the average MM on Fe atoms is equal to $0.31$
$\mu_{B}$; the increase of concentration up to 50\% results in the
increase of MM up to $0.67$$\mu_{B}$ . Notice that at the equilibrium
lattice parameters of pure $\alpha-\mathrm{FeSi}{}_{2}$ but without
structure optimization by the atom coordinates the MM on Fe atoms
does not arise even at 100\% concentration of Li atom. This proves
that emergence of the magnetism in intercalated $\alpha-\mathrm{FeSi}{}_{2}$ primarly
associated with the lattice distortions.

Notice that in order to obtain a sample with large spin polarization,
the optimal concentration of Li intercalate has to be found. Indeed,
since a move along Li concentration from $x_{Li}=0$ towards $x_{Li}=1$
induces magnetism, the corresponding $e_{g}^{\shortuparrow}$ -peak
in the DOS moves from the region above the Fermi level $E_{f}$ at
$x_{Li}=0$ to the region below it at $x_{Li}=1$, while $e_{g}^{\downarrow}$
- peak remains above $E_{f}$ . In a certain range of the concentration
the $e_{g}^{\shortuparrow}$ -peak passes through $E_{f}$ (see Fig.10).
In our case such concentration is in the vicinity of $x_{Li}=0:25.$
The latter provides large spin polarization. This conclusion is obtained
for fully optimized structures.

\begin{figure}
\includegraphics[scale=0.5]{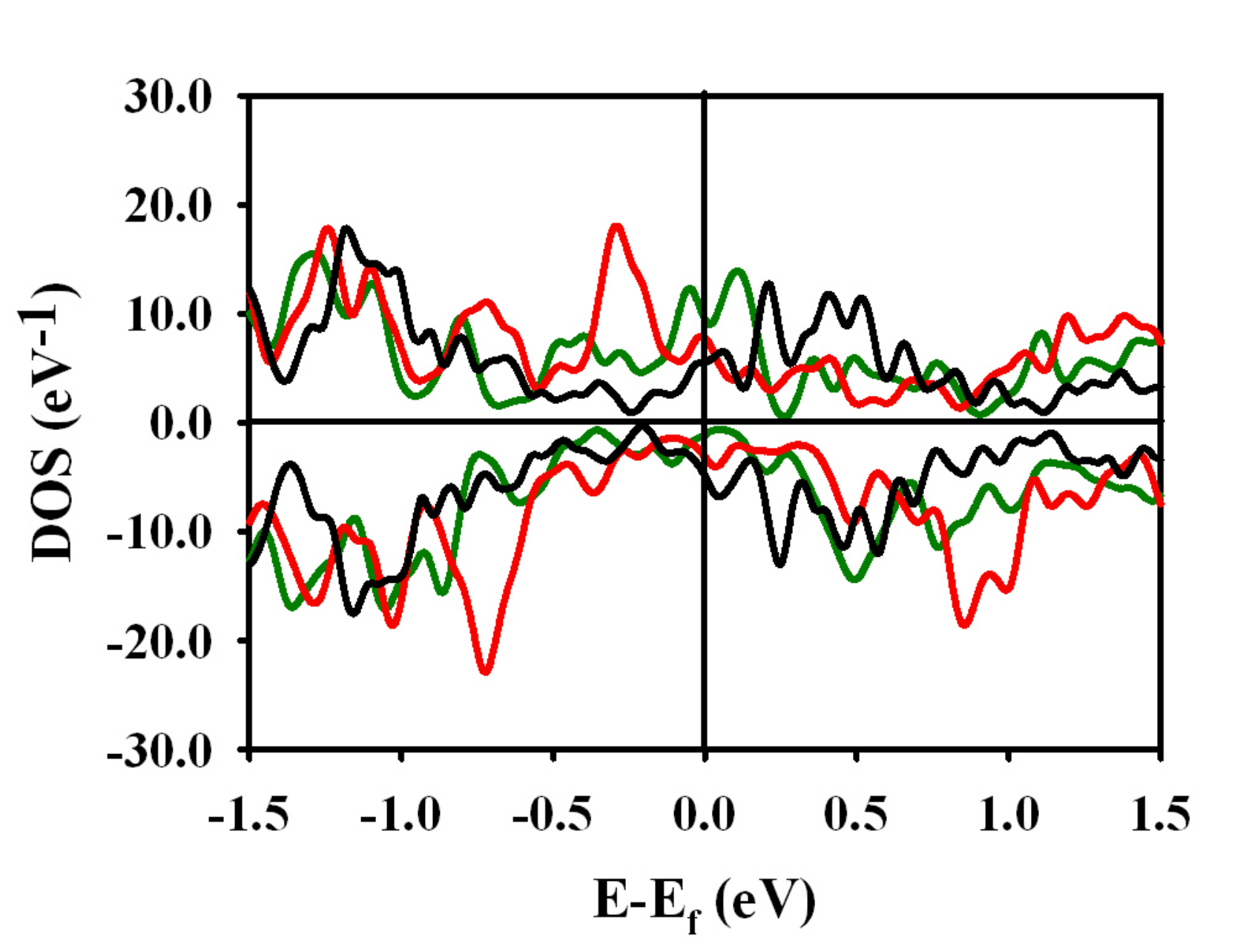}\caption{Full DOS of $Li$-intercalated $\alpha-\mathrm{FeSi}{}_{2}$ for 0\%
(black line), 25\% (green line) and 50\% (red line) concentrations
of intercalated atoms. Zero on the energy axis is the Fermi energy.
\label{fig:Full-DOS_on_x}}

\end{figure}

When we intercalate $\alpha-\mathrm{FeSi}{}_{2}$ by non-metal oxygen
atoms the value of magnetic moment decreases with decrease of oxygen
concentration. However the spin polarization practically does not change
with concentration. Fig. \ref{fig:DOSwIntercalation} shows that the
positions of the $d$-electron peaks in DOS are much more sensitive
to the intercalation by the heavier atoms (such as antimony), than
by atoms of a metal. The intercalation by light oxygen or phosphorus
atoms shifts the same $d$-electron peak with increase of concentration
much less: at 100\% concentration the peak is shifted by $0.25$eV
reaching the Fermi level.

\begin{figure}
\begin{tabular}{cc}
\includegraphics[scale=0.4]{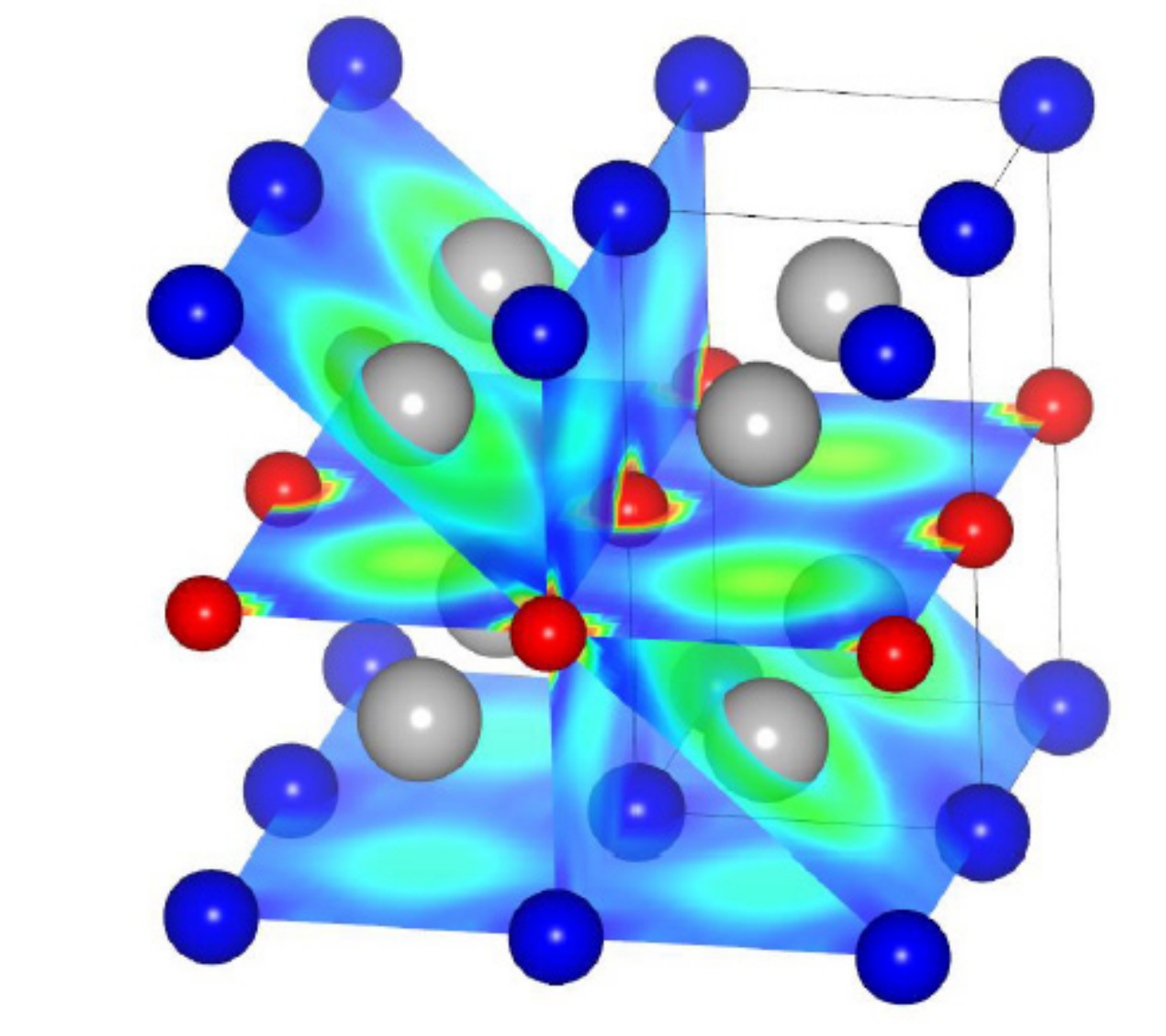} & \includegraphics[scale=0.4]{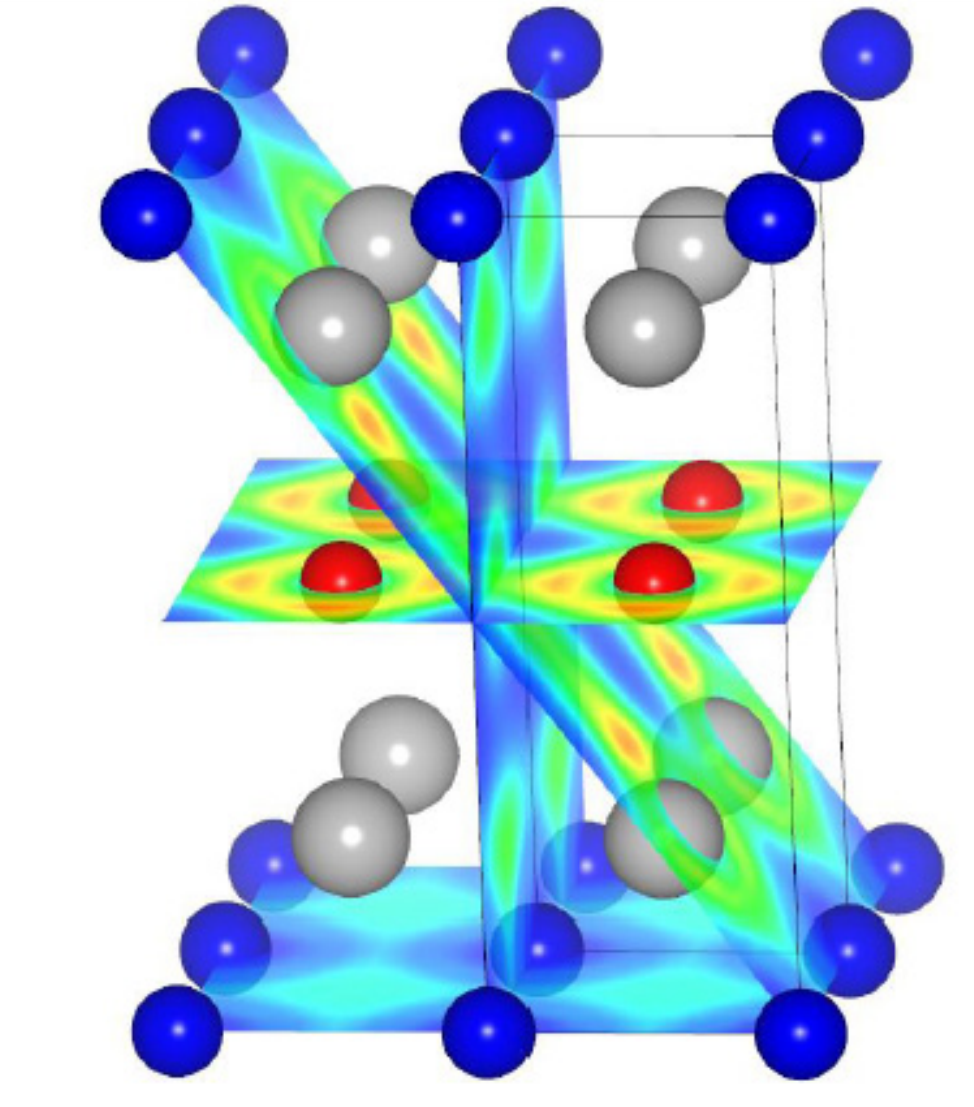}\tabularnewline
a & b\tabularnewline
\end{tabular}\caption{Color online. The electron localization function (ELF) for (a) Li-intercalated
$\alpha-\mathrm{FeSi}{}_{2}$ ; (b) P-intercalated $\alpha-\mathrm{FeSi}{}_{2}$
Blue and green colours correspond to the delocalized electrons, yellow
and red colours display the localized electrons. Blue balls represent
the Fe atoms, grey balls stand for Si atoms, red balls represent intercalated
atoms.\label{fig:ELF}}
\end{figure}

Since there are preferable positions for the metal and non-metal intercalates,
one may expect that the intercalation may cause an anisotropy of the
compound properties. One of ways would be to inspect the tensor of
static electroconductivity $\sigma_{0}$, which in the VASP package
is calculated by means of the Drude formula $\sigma_{\alpha\beta}=\tau\Omega_{\alpha\beta}^{2}/\left(4\pi\right)$.
Here $\Omega_{\alpha\beta}$ is the plasma frequency and $\tau$ is
the relaxation time. However, $\tau$ is the parameter which depends
on many factors (like, e.g., the way of preparation of the sample)
and may differ for different samples even with the same concentration
of the intercalates, not to speak of compounds with different intercalates.
For this reason we prefer to estimate the degree of anisotropy of
a compound just from the ratio $\eta=\Omega_{xx}/\Omega_{zz}$ (notice
that $\Omega_{xx}=\Omega_{yy}).$ The results of calculations are
shown in Table \ref{tab:Tab1}, in 8-th column. A question arises
if the different preferable positions for the metal and non-metal
atoms in lattice of $\alpha-\mathrm{FeSi}{}_{2}$ can be associated
with the anisotropy? We inspected the maps of electronic localization
function (ELF) at the intercalation by Li and P atoms (Fig. \ref{fig:ELF}).
For the non-metal atoms, which prefer to locate on the Si - Si bond,
the conductive channels in $xy$ plane are retained. In turn, the
metal intercalated atoms prefer the position on the bond between out-plane
Fe - Fe atoms. This leads to overlap of conductive channels. The latter
leads to more uniform distribution of delocalized electrons by volume
of crystal (Fig. 11). Although this difference seems to exist for
all metal and non-metal intercalated compounds, its contribution to
anisotropy is not monotonic and a general rule does not exist.

\section{Conclusion}

\textcolor{black}{The fact that the large, if not the decisive, role
in the mechanism of the magnetic structure formation in different
compounds is played by the local environment of the magnetic species
is well-known from the physics of surface and interfaces. In earlier
works \cite{key-10,key-12} in the framework of suggested by us approach
(hybrid self-consistent mapping approach (HSCMA)) we have shown that
the distortions of crystal lattice have a significant impact on the
magnetic moment formation along with types of atoms in the local environment.}
According to the latest experimental data, $\alpha-\mathrm{FeSi}{}_{2}$
is predisposed to the appearance of ferromagnetism in it.\textcolor{black}{{}
In Ref. \cite{key-10} we studied the possible reasons for this. As
it follows from the analysis of t}he map of the magnetic moment dependencies
on the hopping integrals,\textcolor{black}{{} the} hopping integral,\textcolor{black}{{}
which is responsible for the distortion of the crystal lattice in
the $Fe$ plane, plays a crucial role in the appearance of ferromagnetism.
The distinctive feature of all the calculated maps is the presence
of sharp boundaries between region with magnetic states and non-magnetic
ones}. Therefore, the system is in the vicinity of magnetic instability
and it is reasonable to assume that some other type of crystal-lattice
distortions can cause the formation of magnetic state in $\alpha-\mathrm{FeSi}{}_{2}$

In the present work we consider the conditions which can lead to the
appearance of magnetic state in $\alpha-\mathrm{FeSi}{}_{2}$. As
it follows from our analysis within a model, the magnetic state can
arise not only when the distance between in-plane Fe - Fe atoms is
changed, but also, for example, when the distance between out-of-plane
Si - Si atoms is changed. Unfortunately, pronounced magnetic moment
can arise only at quite large distortions of the crystal lattice,
even if a comlex set of distortions is applied. We suggest that intercalation
of $\alpha-\mathrm{FeSi}{}_{2}$ could be a way to solve this problem.
Actually, our calculations show that the intercalation of $\alpha-\mathrm{FeSi}{}_{2}$
results in the appearance of significant magnetic moment on Fe atoms
(0.5 - 1$\mu_{B}$) at the relatively small lattice distortions (Table
\ref{tab:Tab1}). Notice, that it is hardly possible to reproduce
the complex set of lattice distortions caused by the intercalated
atoms by selecting different substrates for the film or nanoparticle
fabrication. Besides, we expect the appearance of a large spin polarization
(60-80\%) in  $\alpha-\mathrm{FeSi}{}_{2}$ intercalated by some atoms,
such as Li, P, Na, O, which cause a reconstruction of the electronic
structure. Such high value of spin polarization makes intercalated
$\alpha-\mathrm{FeSi}{}_{2}$ a promising candidate for the application
in spintronics. Although it is hardly possible to achieve 100\% concentration
of intercalated atoms in practice, the high spin polarization remains
large even at smaller concentrations. Moreover, maintaining a magnetic
state at the smaller concentration of intercalated atoms, the small
distortions of lattice allow experimental fabrication of films on
the silicon substrate, which is extremely important for the modern
silicon technology.
\begin{acknowledgments}
This work was supported by the Russian Fund for Basic Research, Government
of Krasnoyarsk Territory and Krasnoyarsk Region Science and Technology
Support Fund to the research Projects No 17-42-240212 and No 18-42-243019.\end{acknowledgments}

\end{document}